\documentclass[notitlepage,superscriptaddress,nofootinbib,amsmath,amssymb,aps,11pt]{revtex4-1}
\usepackage[]{inputenc}
\usepackage[dvips]{graphicx}
\usepackage{color}
\usepackage{bm}
\usepackage{mathpazo}
\usepackage{epstopdf}
\usepackage[bookmarks=true,colorlinks,linkcolor=blue,urlcolor=blue,citecolor=red]{hyperref}
\renewcommand\a{\alpha}
\renewcommand\b{\beta}
\renewcommand\d{\delta}

\renewcommand\l{\lambda}
\renewcommand\r{\rho}

\renewcommand\t{\tau}

\renewcommand\c{\chi}
\renewcommand\j{\psi}
\renewcommand\o{\omega}
\newcommand\e{\epsilon}
\newcommand\g{\gamma}
\newcommand\z{\zeta}
\newcommand\m{\mu}
\newcommand\n{\nu}

\newcommand\p{\pi}
\newcommand\h{\theta}
\newcommand\s{\sigma}
\newcommand\f{\phi}
\newcommand\w{\eta}

\newcommand\vf{\varphi}
\renewcommand\L{\Lambda}
\renewcommand\P{\Pi}
\renewcommand\S{\Sigma}
\renewcommand\O{\Omega}

\newcommand\D{\Delta}
\newcommand\G{\Gamma}

\newcommand\J{\Psi}

\newcommand{\fig}[1]{Fig.~\ref{#1}}
\newcommand{\eq}[1]{Eq.~(\ref{#1})}
\newcommand{\sect}[1]{Sec.~\ref{#1}}
\newcommand{\eqs}[2]{Eqs.~(\ref{#1})-(\ref{#2})}

\newcommand\lb{\left(}
\newcommand\rb{\right)}
\newcommand\ls{\left[}
\newcommand\rs{\right]}
\newcommand\lc{\left\{}
\newcommand\rc{\right\}}
\newcommand{\lan}{\langle}
\newcommand{\ran}{\rangle}

\newcommand\ra{\rightarrow}

\newcommand{\non}{\nonumber\\}
\newcommand\pt{\partial}

\newcommand{\diag}{{\rm{diag}}}

\newcommand{\Tr}{{\rm Tr}}

\newcommand{\bx}{{\vec x}}

\newcommand{\bp}{{\vec p}}

\newcommand{\bB}{{\vec B}}
\newcommand{\bE}{{\vec E}}
\newcommand{\bJ}{{\vec J}}

\newcommand{\jb}{{\bar \j}}

\newcommand{\rp}{{\rm RP}}

\renewcommand{\part}{{\rm part}}

\usepackage{slashed}
\usepackage{mathrsfs}

\newcommand\mc{\mathcal}
\newcommand\ms{\mathscr}
\newcommand\na{\nabla}
\newcommand\ola{\overleftarrow}

\newcommand{\hzero}{{\hat{0}}}

\renewcommand{\vec}{\boldsymbol}
\newcommand\be{\begin{equation}}
\newcommand\ba{\begin{eqnarray}}
\newcommand\ee{\end{equation}}
\newcommand\ea{\end{eqnarray}}

\begin{document}

\title{Anomalous chiral transports and spin polarization in heavy-ion collisions}
\author{\normalsize{Yu-Chen Liu}}
\affiliation{Physics Department and Center for Particle Physics and Field Theory, Fudan University, Shanghai 200433, China.}
\author{\normalsize{Xu-Guang Huang}}
\affiliation{Physics Department and Center for Particle Physics and Field Theory, Fudan University, Shanghai 200433, China.}
\affiliation{Key Laboratory of Nuclear Physics and Ion-beam Application (MOE), Fudan University, Shanghai 200433, China.}

\begin{abstract}
Relativistic heavy-ion collisions create hot quark-gluon plasma as well as very strong electromagnetic (EM) and fluid vortical fields. The strong EM field and vorticity can induce intriguing macroscopic quantum phenomena such as chiral magnetic, chiral separation, chiral electric separation, and chiral vortical effects as well as the spin polarization of hadrons. These phenomena provide us with experimentally feasible means to study the nontrivial topological sector of quantum chromodynamics, the possible parity violation of strong interaction at high temperature, and the subatomic spintronics of quark-gluon plasma. These studies, both in theory and in experiments, are strongly connected with other subfields of physics such as condensed matter physics, astrophysics, and cold atomic physics, and thus form an emerging interdisciplinary research area. We give an introduction to the aforementioned phenomena induced by the EM field and vorticity and an overview of the current status of their experimental research in heavy-ion collisions. We also briefly discuss spin hydrodynamics as well as chiral and spin kinetic theories.
\end{abstract}

\maketitle

\section {Introduction}\label{sec:intro}
It is well known that the strong interaction binds quarks and gluons together to form hadrons such as protons and neutrons. The contemporary theory of strong interaction is governed by quantum chromodynamics (QCD), which is an $SU(3)$ quantum gauge theory. The non-Abelian nature of QCD has important consequences such as color confinement at a low-energy scale and asymptotic freedom at a high-energy scale. Color confinement means that at low-energy scales, the color carriers (i.e., quarks and gluons) are always confined in color singlet hadrons; thus, no isolated quark and gluon can be observed. However, when the energy scale grows (e.g., when the temperature or the baryon density of the hadronic matter is increased), QCD undergoes a deconfinement phase transition, and quarks and gluons are liberated from the hadrons. When the energy scale is very high, the coupling constant of QCD becomes small and the system goes into the perturbative regime of QCD. In this regime, the coupling constant decreases with increasing energy scale, a phenomenon known as asymptotic freedom. Reliable perturbative calculation can apply in this regime.

In reality, the conditions for the deconfinement phase transition are difficult to achieve. Moreover, the confinement energy scale of QCD is approximately $\Lambda_{\rm QCD}\sim 200$ MeV, which, in terms of temperature, is approximately $T_c\sim\Lambda_{\rm QCD}\sim 10^{12}$ K. This high temperature may have once existed in the early universe (e.g., according to modern cosmology, this occurred immediately following the Big Bang) and can currently only be realized experimentally on earth by relativistic heavy-ion collisions. Current operating facilities of heavy-ion collisions include the Relativistic Heavy-Ion Collider (RHIC) at Brookhaven National Laboratory in the United States of America and the Large Hadron Collider (LHC) at the European Organization for Nuclear Research (CERN). RHIC has been operational since 2000 and its current top colliding energy for Au + Au collisions is $\sqrt{s}=200$ GeV. LHC has been in operation since 2010 and its current top colliding energy for Pb + Pb collisions is $\sqrt{s}=5.02$ TeV. In these colliders, two counterpropagating beams of ions are accelerated to ultrahigh speed to make them collide. The large kinematic energies of the ions accumulate at the colliding point so that the transient energy density can be sufficiently high to achieve the deconfinement phase transition. The deconfined quark-gluon matter produced from this phase transition is typically known as quark-gluon plasma (QGP). The data collected at RHIC and LHC have indicated strong evidence of the existence of QGP and also revealed numerous extraordinary properties of QGP. Here, we list a few; more discussions can be found in Ref.~\cite{qgp:wang}. The QGP is considered to be the ``most perfect fluid" because the ratio of its shear viscosity to its entropy density is the smallest among those of all the known fluids, including the helium superfluid. The QGP can strongly quench the energetic jets (i.e., a particle or a collimated shower of particles of high transverse momenta), a phenomenon known as jet quenching, which indicates that the energetic jets interact strongly with the constituents of QGP. The color force between two heavy quarks may be screened in QGP, similar to the usual Debye screening of the electric charges in electromagnetic (EM) plasmas. This enables heavy quarkonia, such as the $J/\J$, to be easily dissociated in QGP, leading to a suppression in the final measured yields.

In addition to the abovementioned phenomena, in recent years, researchers have realized that relativistic heavy-ion collisions can also generate strong EM fields and fluid vorticity. More importantly, under these strong EM fields and vorticity, numerous intriguing macroscopic quantum phenomena may occur. These phenomena provide us opportunities to study the nontrivial chiral properties of quark-gluon matter, particularly those related to quantum anomaly, as well as the spin dynamics of QGP. Moreover, these phenomena are closely related to other subfields of physics, such as particle physics, condensed matter physics, astrophysics, and cold atomic physics, and thus give rise to a new interdisciplinary research area. Some review articles are already available, including Refs.~\cite{Kharzeev:2015znc,Huang:2015oca,Hattori:2016emy,Skokov:2016yrj,Zhao:2019hta,Li:2020dwr,Becattini:2020ngo,Huang:2020xyr}. In the following section, we introduce the EM field and vorticity that occur in heavy-ion collisions.

\section {EM field and vorticity}\label{sec:fields}
\begin{figure}[!t]
\begin{center}
\includegraphics[width=7.0cm]{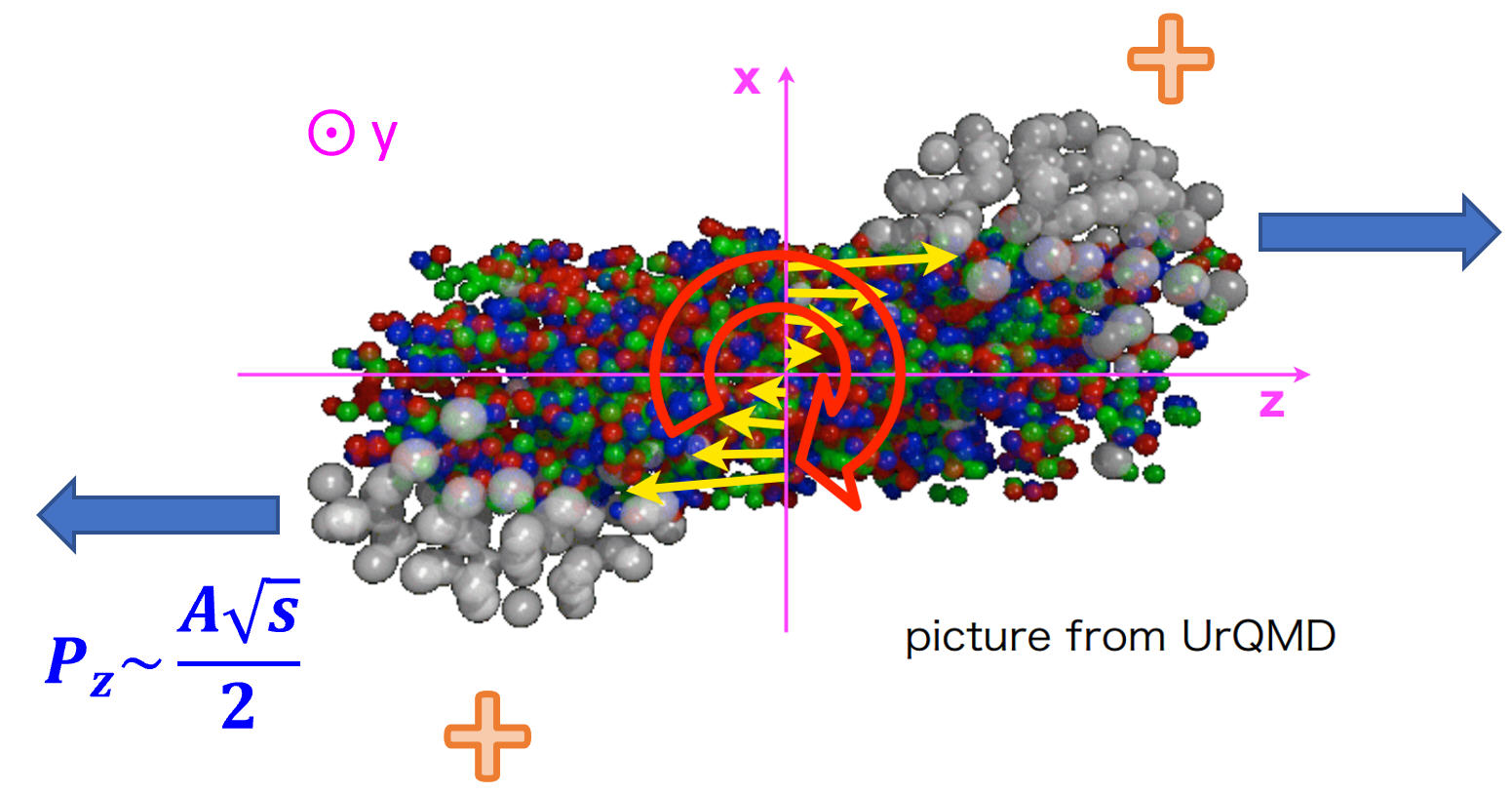}
\caption{Geometry of a typical noncentral collision. The figure was modified from https://urqmd.org.}
\label{illus}
\end{center}
\end{figure}

Let us consider a noncentral collision between two nuclei. The collision geometry is depicted in \fig{illus}. The $z$ direction is along the motion of the projectile, the $x$ direction is along the impact parameter $\bm{b}$ (from the target to the projectile), and the $y$ direction is along $\hat{\bm{z}}\times\hat{\vec x}$. The $x$-$z$ plane is the reaction plane. As the nucleus is positively charged, its motion generates an electric current that generates a magnetic field. At the moment of collision, because of geometric symmetry, a magnetic field perpendicular to the reaction plane is produced at the collision center ($\bx=\bm 0$). Let us estimate the strength of this magnetic field by using the Biot--Savart formula. For a Au + Au collision at $\sqrt{s}=200$ GeV with $b=10$ fm, we have
\begin{eqnarray}
\label{estimate}
eB_y\approx -2Z_{\rm Au}\g\frac{e^2}{4\p}\frac{v_z}{(b/2)^2}\approx -10 m_\p^2\approx -10^{19}\;{\rm Gauss},
\end{eqnarray}
where $v_z=\sqrt{1-(2m_N/\sqrt{s})^2}\approx 0.99995$ is the velocity of the nucleus in the laboratory frame in which $m_N$ is the nucleon mass, $\g=1/\sqrt{1-v_z^2}\approx 100$ is the Lorentz factor, and $Z_{\rm Au}=79$ is the proton number of the Au nucleus.

\begin{figure}[!t]
\begin{center}
\includegraphics[width=7.0cm]{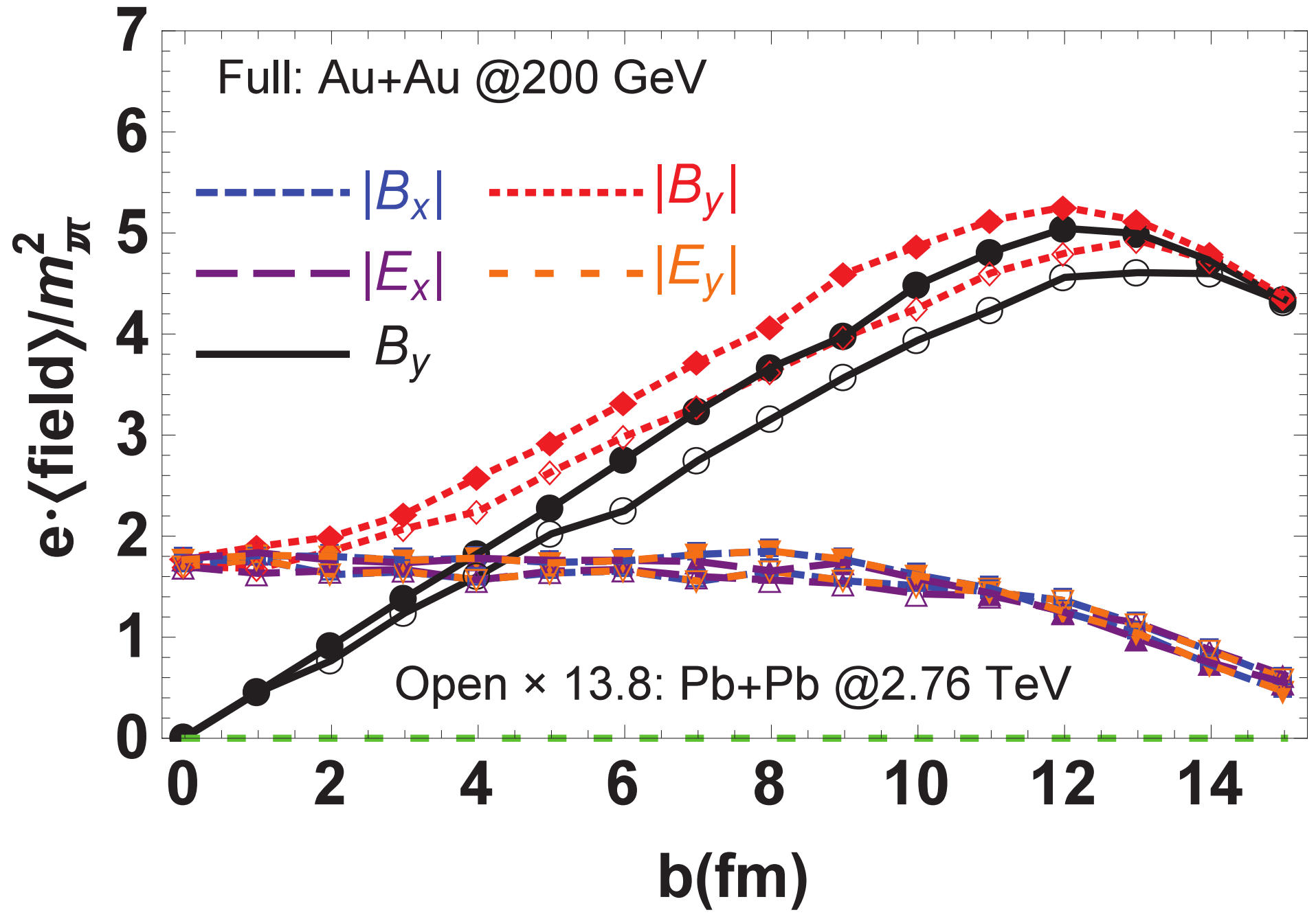}
\caption{EM field versus the impact parameter in heavy-ion collisions (note that $m_\p^2/e\approx 3.3\times10^{18}$ Gauss). The figure is reproduced from Ref.~\cite{Deng:2012pc}.}
\label{emfield}
\end{center}
\end{figure}

This is a huge magnetic field, considerably larger than the squared masses of the electron and light quarks ($u, d$ quarks), and thus may induce significant quantum effects in systems composed of electrons and light quarks. Moreover, this is the strongest known magnetic field in the current universe; it is several orders stronger than the surface magnetic fields of neutron stars, including magnetars ($eB\sim10^{14}-10^{15}$ Gauss)~\cite{Kaspi:2017fwg}. The result in \eq{estimate} is very rough. More advanced simulations can be performed using transport models such as HIJING, AMPT, UrQMD~\cite{Skokov:2009qp,Voronyuk:2011jd,Bzdak:2011yy,Ou:2011fm,Deng:2012pc,Bloczynski:2012en,Bloczynski:2013mca,Deng:2014uja,Zhong:2014cda,Zhong:2014sua,Huang:2015fqj,Deng:2016knn,Huang:2017azw,Zhao:2017rpf,Deng:2017ljz,Cheng:2019qsn}. In such simulations, one can determine the positions and momenta of each charged particle before and after the collision and then use, for example, the Lienard--Wiechert formula to calculate the EM fields. The possible quantum correction to the Lienard--Wiechert formula can be estimated (which was found to be insignificant)~\cite{Bloczynski:2012en,Huang:2015oca}. Many aspects of the EM field were studied through this approach, such as the event-by-event fluctuations of the strength and orientation of the EM fields~\cite{Bzdak:2011yy,Deng:2012pc,Bloczynski:2012en}, azimuthal correlation between the EM field and matter geometry~\cite{Bloczynski:2012en,Bloczynski:2013mca}, EM fields in different collision systems~\cite{Deng:2014uja,Deng:2016knn}, and influence of the charge distribution of nucleons~\cite{Bloczynski:2012en,Bloczynski:2013mca}  (please see the reviews ~\cite{Huang:2015oca,Hattori:2016emy}). In \fig{emfield}, we show the impact parameter dependence of the EM fields computed using the HIJING model for Au + Au and Pb + Pb collisions at RHIC and LHC energies, respectively. It is seen that the strength of the fields is roughly proportional to the collision energy $\sqrt{s}$~\cite{Deng:2012pc}.

Let us consider once again a noncentral collision of energy $\sqrt{s}$ and impact parameter $b$. The system possesses an angular momentum
\begin{eqnarray}
J_y\approx -\frac{Ab\sqrt{s}}{2},
\end{eqnarray}
where $A$ is the mass number of the nucleus. For RHIC Au + Au collisions at $\sqrt{s}=200$ GeV and $b=10$ fm, we obtain $|J_y|\approx 10^6\hbar$. Compared to the total spin of the produced hadrons (e.g., for a typical number of produced hadrons of $1000$, the total spin would be $\sim 10^3\hbar$),this is a large angular momentum. After the collision, a part of this angular momentum is transferred to the produced QGP. As the equation of state of the QGP is very soft, this part of the angular momentum does not cause the rigid rotation of the QGP but rather induces local fluid vortices. The strength of a fluid vortex is described by the vorticity. In nonrelativistic hydrodynamics, the vorticity is defined by
\begin{eqnarray}
\bm{\o}=\frac{1}{2}\bm\nabla\times\bm v,
\end{eqnarray}
where $\bm v$ is the flow velocity. From this definition, it is clear that the physical meaning of the vorticity is the local angular velocity of the fluid cell. In relativistic hydrodynamics, according to different physical contexts, different vorticities can be defined. The commonly used ones are the kinematic, temperature, and thermal vorticities. The kinematic vorticity is a natural generalization of the nonrelativistic vorticity:
\begin{eqnarray}
\o^\m=\frac{1}{2}\e^{\m\n\r\s}u_\n\pt_\r u_\s,
\end{eqnarray}
where $u^\m=\g(1,\bm v)$ is the flow four velocity. In many situations, it is more convenient to use its tensorial representation $\o_{\m\n}=(1/2)(\pt_\n u_\m-\pt_\m u_\n)$, which is related to $\o^\m$ by $\o^\m=-(1/2)\e^{\m\n\r\s}u_\n\o_{\r\s}$. The temperature vorticity is defined as
\begin{eqnarray}
\o_{\rm T}^{\m}=\frac{1}{2}\e^{\m\n\r\s} u_\n\pt_\r (Tu_\s),
\end{eqnarray}
where $T$ is the temperature. The special property of the temperature vorticity is that, for an ideal neutral fluid, it satisfies the Carter--Lichnerowicz equation $\omega^{\rm T}_{\mu\nu}u^\nu=0$, which yields two interesting consequences~\cite{Becattini:2015ska,Deng:2016gyh}. One consequence is the relativistic Helmholtz--Kelvin theorem stating that the flow circulation, defined as $l(\t)=\oint T u_\mu dx^\mu$, is a co-moving invariant of the fluid, $dl/d\t=0$. Another consequence is the conservation of $T\o_{\rm T}^\m$, $\pt_\m(T\o_{\rm T}^{\m})=0$. The conserved charge ${\cal H}_{\rm T}=(1/2)\int d^3\vec x T^2 \gamma^2\vec v\cdot\vec\nabla\times\vec v$ defines the relativistic fluid helicity, which measures the degree of linkage of the vortex lines. The thermal vorticity in tensorial form is defined as
\begin{eqnarray}
\varpi_{\m\n}=\frac{1}{2}[\pt_\n(\b u_\m)-\pt_\m(\b u_\n)],
\end{eqnarray}
where $\b=1/T$ is the inverse temperature. The importance of thermal vorticity relies on the fact that it characterizes the global equilibrium of a rotating fluid and determines the spin polarization of the constituent particles in the fluid at the global thermal equilibrium~\cite{Becattini:2012tc,Becattini:2013fla}. We will discuss the spin polarization in detail in \sect{sec:sp}.

\begin{figure}[!t]
\begin{center}
\includegraphics[width=7.0cm]{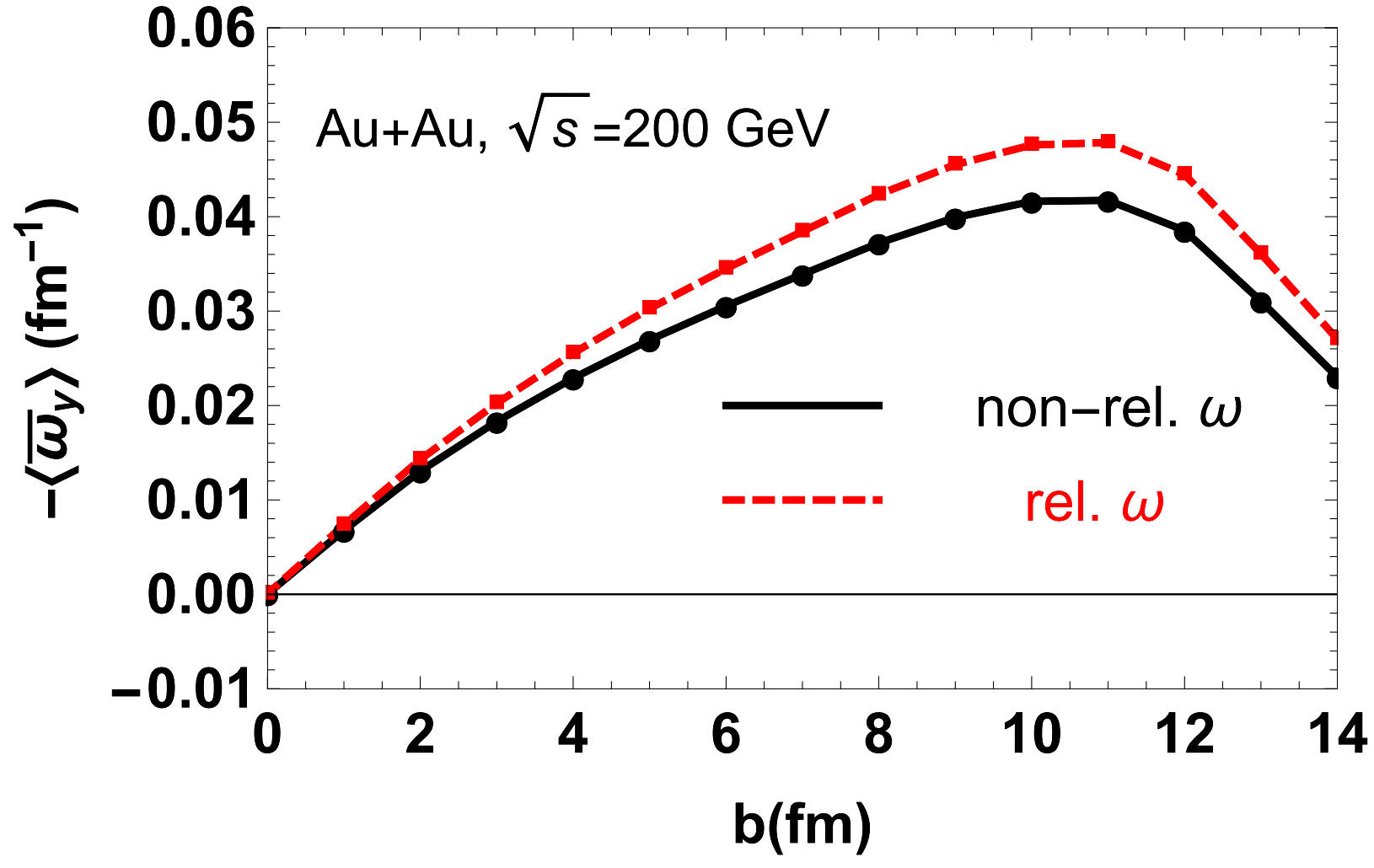}
\caption{Impact parameter dependence of the nonrelativistic and relativistic kinematic vorticities in Au + Au collisions. The figure is from Ref.~\cite{Deng:2016gyh}.}
\label{kin-omg}
\end{center}
\end{figure}
\begin{figure}[!t]
\begin{center}
\includegraphics[width=7.0cm]{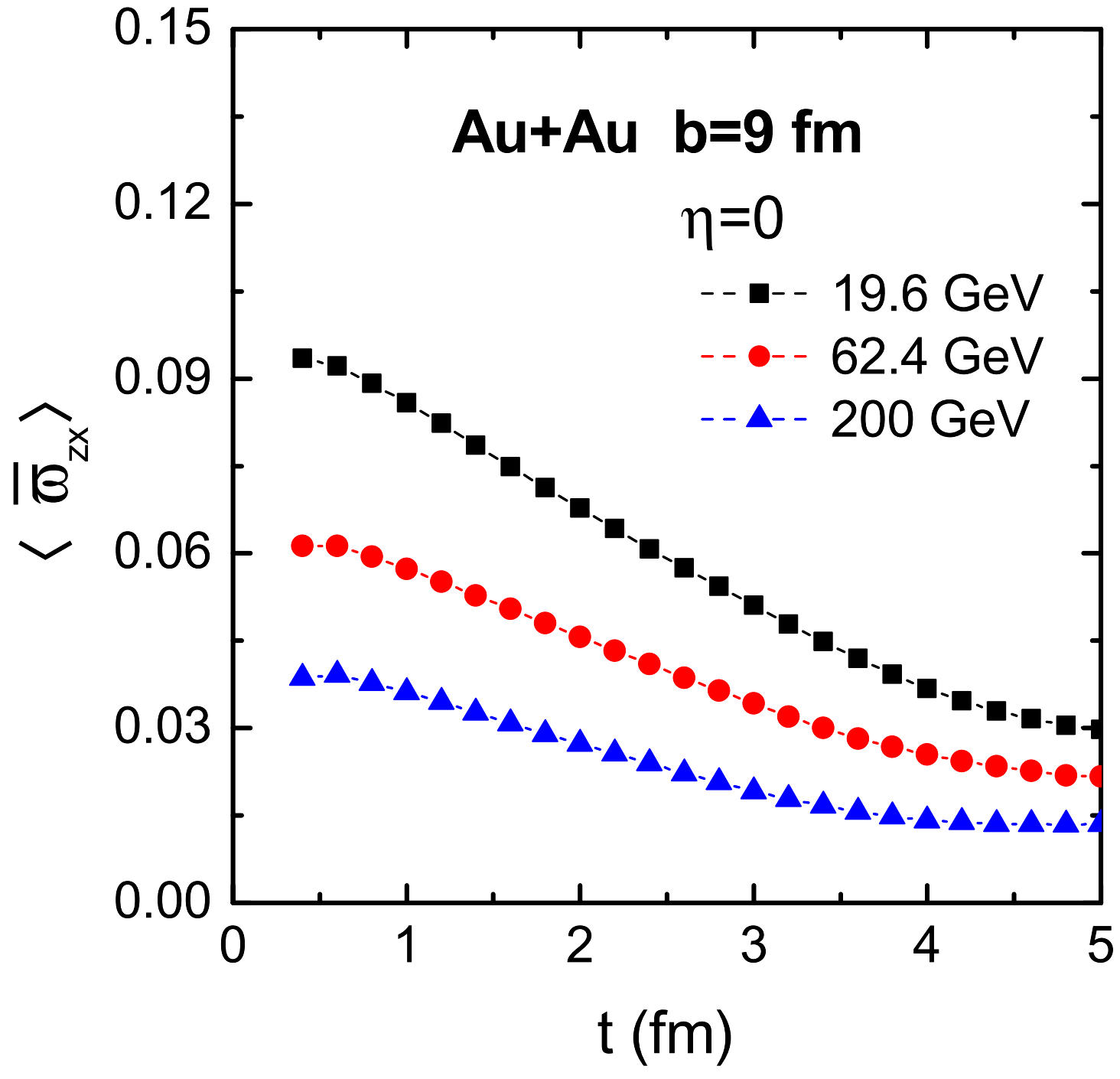}
\caption{Time evolution of the thermal vorticity in Au + Au collisions for several different collision energies. The figure is from Ref.~\cite{Wei:2018zfb}.}
\label{the-omg}
\end{center}
\end{figure}

In \fig{kin-omg}, we present the numerical results of the nonrelativistic and relativistic kinematic vorticities in Au + Au collisions at $\sqrt{s}=200$ GeV based on a HIJING simulation~\cite{Deng:2016gyh}. The results are averaged over the reaction region and over $10^5$ events (please refer Ref~\cite{Deng:2016gyh} for more details). As seen in \fig{kin-omg}, the vorticities grow with $b$ at $b<2 R_A$ (where $R_A$ is the radius of the nucleus) simply because the total angular momentum of the system increases and then decreases at $b\geq 2R_A$ because of the shrinking of the reaction region. Numerical results show that the vorticity can be large (with a peak value of $|\lan\overline{\o}_y\ran|\sim 10$ MeV $\sim 10^{21}$ s$^{-1}$). This is the strongest vorticity we have ever known. For this reason, we sometimes call QGP as the ``most vortical fluid"~\cite{STAR:2017ckg}. In \fig{the-omg}, we show numerical results for the time evolution of the thermal vorticity in Au + Au collisions for $\sqrt{s}=19.6, 62.4,$ and $200$ GeV obtained using the AMPT model~\cite{Wei:2018zfb}. It is natural that the vorticity decays in time because of the fire-ball expansion. However, surprisingly, the vorticity decreases when $\sqrt{s}$ increases; this is a relativistic effect that we will discuss later. The numerical simulations for the vorticities can also be found in Refs.~\cite{Jiang:2016woz,Deng:2016gyh,Wei:2018zfb,Becattini:2015ska,Teryaev:2015gxa,Xie:2016fjj,Ivanov:2017dff,Kolomeitsev:2018svb,Deng:2020ygd}.

\section {Chiral anomaly and transport phenomena}\label{sec:anom}
What are the consequences of strong EM fields and vorticity in heavy-ion collisions? During the past decade, many discussions have addressed this question and considerable interesting effects have been studied. Among the most intriguing effects are the quantum phenomena that are closely related to the spin dynamics of quarks. For massless fermions, these phenomena are also deeply related to the chiral anomaly of QCD and quantum electrodynamics (QED) and can be called anomalous chiral transports (ACTs). For a massive case, the spin polarization of hyperons by vorticity is a remarkable example. Of course, in general, both ACTs and spin polarization could occur with both massless and massive particles, but they manifest mostly with massless and massive particles, respectively. In this section, we focus on ACTs. The noticeable examples of ACTs are the chiral magnetic effect (CME), chiral vortical effects (CVEs), chiral separation effect (CSE), and chiral electric separation effect (CESE). We give a pedagogical discussion of the underlying mechanisms of the ACTs~\cite{Landsteiner:2016led,Huang:2018svw}.
\begin{figure*}
\begin{center}
\includegraphics[width=5.0cm]{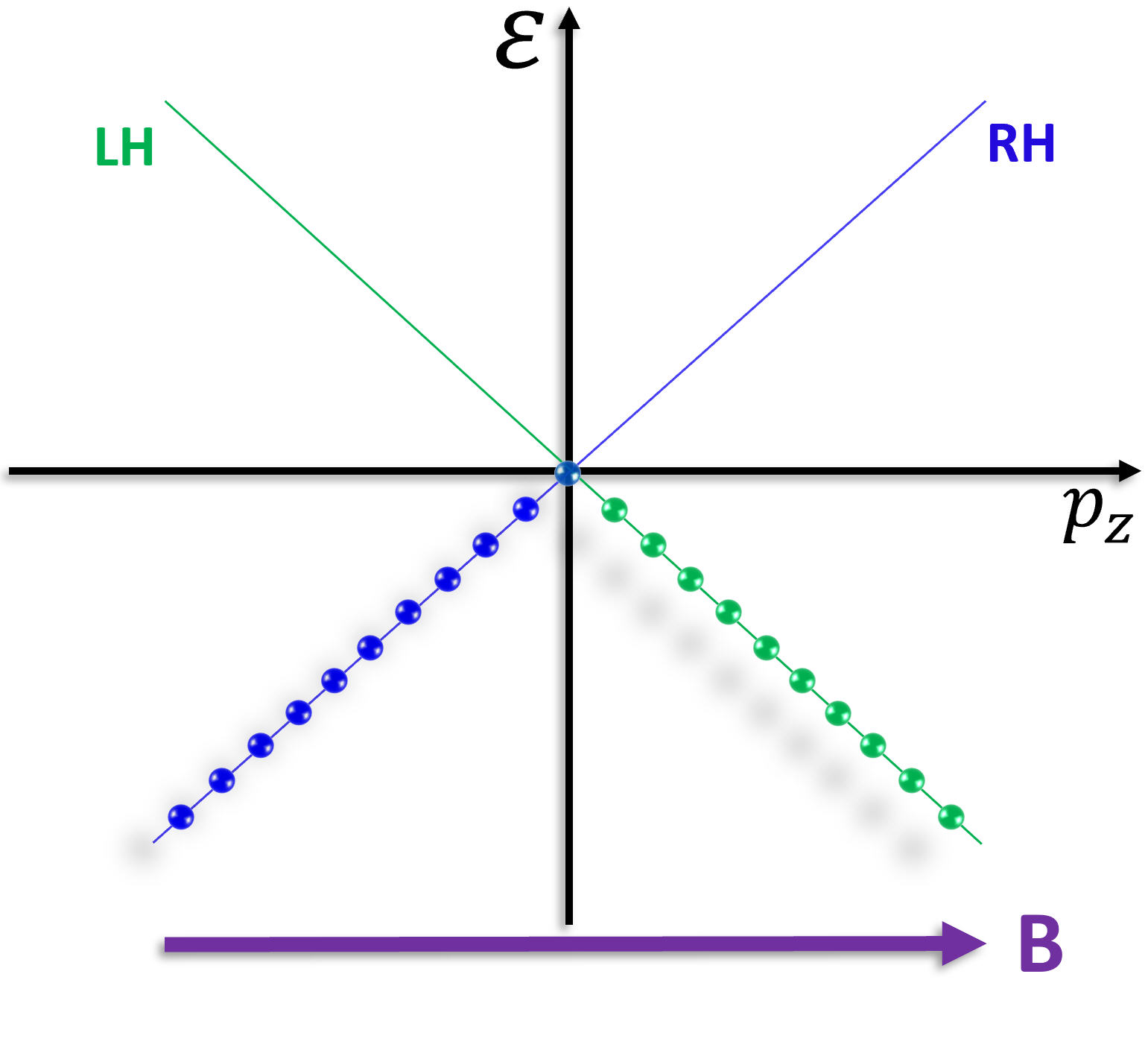}\;\;\;\;\;\;\;\;\;\;\;\;\;\;\;\;
\includegraphics[width=5.0cm]{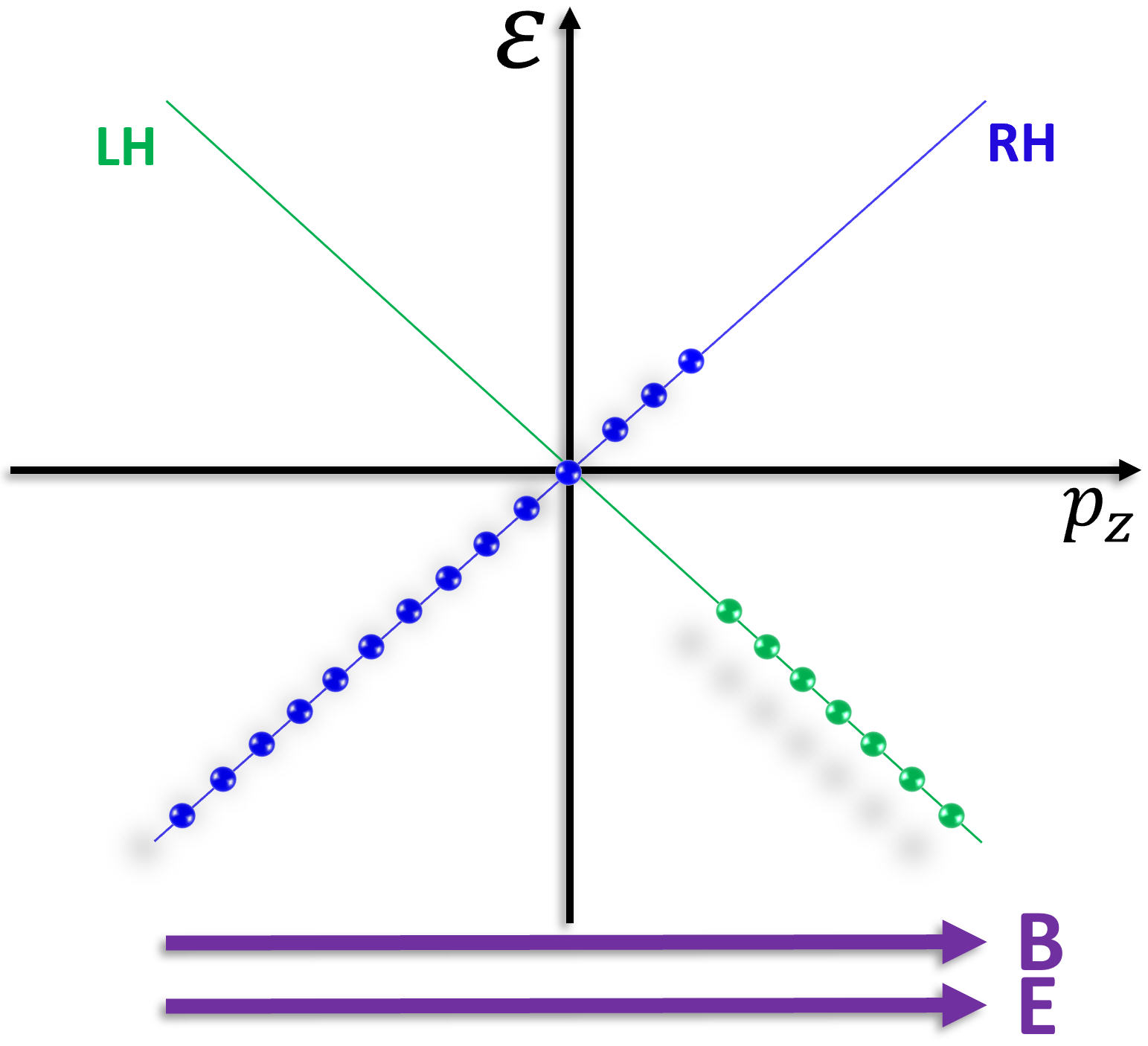}
\caption{(Left) Lowest Landau level in a strong magnetic field. (Right) The electric field induces spectral flow and results in the chiral anomaly.}
\label{fig-ca}
\end{center}
\end{figure*}

Consider a massless Dirac fermion of charge $e>0$ in a strong constant magnetic field along the $z$ direction. This is the usual Landau problem in quantum mechanics. The energy spectrum can be obtained by solving the Dirac equation, and the result is presented as Landau levels,
\begin{eqnarray}
\mathcal{E}_n^2=p_z^2+2neB, \;\; n=0,1,2,\cdots,
\end{eqnarray}
where $n$ labels the Landau levels. The lowest Landau level (LLL), which corresponds to $n=0$, is special; see \fig{fig-ca} (left). First, the LLL is gapless, whereas all the higher Landau levels are gapped by $\sqrt{2neB}$. Thus, for large $eB$, we need to consider only the LLL. Second, the spin of LLL is fully polarized, that is, the LLL is nondegenerate in spin. All the states of the LLL are of spin-up type. In a many-body picture, this means that the LLL fermions are all of spin-up type. Third, the dynamics of the LLL fermion is 1+1 dimensional because the transverse motion is frozen and $\mathcal{E}_{n=0}$ is independent of $B$. We define the chirality for each LLL fermion according to its momentum direction relative to its spin direction. If $p_z$ is parallel to the spin, we call it a right-handed (RH) fermion; if $p_z$ is opposite to its spin, we call it a left-handed (LH) fermion. In this situation, the numbers of RH and LH fermions are conserved separately (i.e., $\pt_\mu J^\mu_{R/L}=0$ with $J^\m_{R/L}=(1/2)\jb\g^\m(1\pm\g_5)\j$, or equivalently $\pt_\mu J^\mu_{V/A}=0$, where the vector and axial currents are defined as $J_{V/A}^\m=J_R^\mu\pm J_L^\mu$).

Now suppose an electric field is imposed in the same direction as the magnetic field; see \fig{fig-ca} (right). Near the level crossing node $p_z=0$, the downward moving particles can be easily flipped by the electric field to move upward, and thus some LH fermions are tuned to RH fermions. This is a typical spectral flow phenomenon. Therefore, $N_V=N_R+N_L$, the total number of RH and LH fermions, is still conserved, whereas the difference $N_A=N_R-N_L$ is not. We can calculate the time derivative of $N_A$ in the following manner. Let $p_F^{R/L}$ denote the Fermi momenta of the RH and LH fermions. We have
\begin{eqnarray}
N_{R/L}=V\frac{p_F^{R/L}}{2\p}\frac{eB}{2\p},
\end{eqnarray}
where $eB/(2\p)$ is the transverse density of state and $V$ is the volume of the system. The electric force gives $\dot{p}_F^{R/L}=\pm eE$. Thus,
\begin{eqnarray}
\frac{dN_{R/L}}{dt}=V\frac{\dot{p}_F^{R/L}}{2\p}\frac{eB}{2\p}=\pm V\frac{eE}{2\p}\frac{eB}{2\p},
\end{eqnarray}
or equivalently, $d N_V/dt=0$ and $d N_A/dt=Ve^2EB/(2\p^2)$. In differential forms, they yield $\pt_\m J_V^\m=0$ and
\begin{eqnarray}
\label{chiralanom}
\pt_\mu J_A^\m=\frac{e^2}{2\p^2}\bE\cdot \bB.
\end{eqnarray}
This is the well-known chiral or axial anomaly~\cite{Adler:1969gk,Bell:1969ts}. We note that although we obtain \eq{chiralanom} by considering the strong magnetic field so that only the LLL is occupied, the result is actually true for an arbitrary magnetic field, as the higher Landau levels are degenerate in chirality and do not contribute to \eq{chiralanom}.
\begin{figure*}
\begin{center}
\includegraphics[width=6cm]{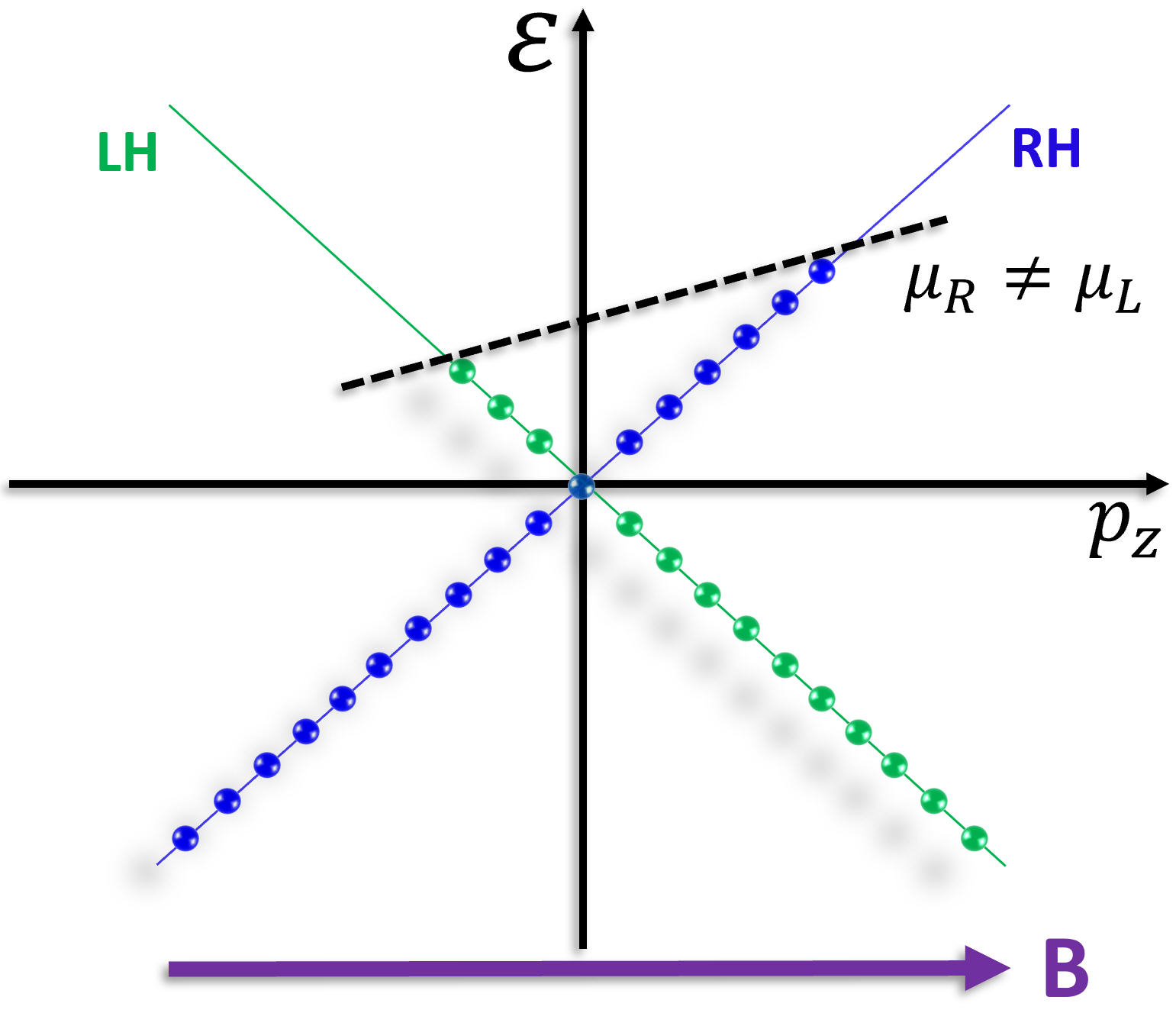}
\caption{Emergence of the chiral magnetic and separation effects.}
\label{fig-ckt}
\end{center}
\end{figure*}

With the previous preparation, we now remove the electric field and calculate the RH and LH currents along the magnetic field; see \fig{fig-ckt}. A current is equal to the carrier density times the velocity of the constituent particles. For massless particles, the velocity is the speed of light such that
\begin{eqnarray}
\label{cmecur1}
J_{R/L}=\pm n_{R/L}=\pm\frac{p_F^{R/L}}{2\p}\frac{eB}{2\p},
\end{eqnarray}
where the minus sign is because that LH fermions move opposite to the direction of the magnetic field. We can rewrite \eq{cmecur1} as
\begin{eqnarray}
\label{cmecur2}
J_{V}=\frac{p_F^{R}-p_F^L}{2\p}\frac{eB}{2\p}=\frac{\m_A}{2\p^2}eB,
\end{eqnarray}
and
\begin{eqnarray}
\label{csecu}
J_{A}=\frac{p_F^{R}+p_F^L}{2\p}\frac{eB}{2\p}=\frac{\m_V}{2\p^2}eB,
\end{eqnarray}
where we have defined the vector and axial chemical potentials as $\m_{V/A}=(p_F^R\pm p_F^L)/2$. The current (\ref{cmecur2}) is the CME current~\cite{Kharzeev:2007jp,Fukushima:2008xe}, and the current (\ref{csecu}) is the CSE current~\cite{Son:2004tq,Metlitski:2005pr}, which appears even when $p_F^R=p_F^L$. The CME exhibits very special properties. First, it is a macroscopic quantum effect. Second, its occurrence requires P and CP violations in the medium. Third, the generation of the CME current is time-reversal even, i.e.,  no associated entropy production occurs. Thus, the CME current is a type of superconducting current. We also must emphasize that the CME conductivity is fixed by the chiral anomaly and is thus free of renormalization.

In classical physics, the Larmor theorem establishes that the motion of a charged particle of mass $m$ in a magnetic field is equivalent to the motion in a rotating frame with frequency $eB/(2m)$. This suggests the existence of analogous effects to CME and CSE but induced by rotation or vorticity. Consider a massless particle in a rotating frame. The particle feels a Coriolis force $\vec F=2p \dot{\vec x}\times\vec\o + O(\o^2)$, where $\vec\o$ is the rotating frequency. We have assumed that $\o$ is so small that we neglect the centrifugal force, which is $O(\o^2)$. As the Coriolis force is very similar to the Lorentz force (replacing $eB$ by $2p\o$), we can consider the ``Landau level problem'' in the rotating frame. Let us again consider only the LLL and consider a many-body system co-rotating with the frame. Compared to the magnetic case, the only difference here is that the expression for the density is modified: $n_{R/L}=(2\p)^{-2}\int_0^{p_F^{R/L}}dp_z 2p_z\o =(p_F^{R/L})^2\o/(2\p)^2$. Now the currents read
\begin{eqnarray}
\label{cvecu1}
J_{V}&=&n_R-n_L=\frac{\m_V\m_A}{\p^2}\o,\\
\label{cvecu2}
J_{A}&=&n_R+n_L=\frac{\m_V^2+\m_A^2}{2\p^2}\o.
\end{eqnarray}
These are the vector and axial CVEs~\cite{Vilenkin:1979ui,Erdmenger:2008rm,Banerjee:2008th,Son:2009tf}. A more rigorous consideration shows that an additional term, namely, $T^2\o/6$ exists in $J_A$, which may be related to the global gravitational anomaly~\cite{Landsteiner:2011cp,Glorioso:2017lcn}.

In \fig{emfield}, we see that, in addition to the strong magnetic field, heavy-ion collisions also create a strong electric field because of the fluctuation of the proton distribution. In geometrically asymmetric collisions such as Cu + Au collisions, a strong electric field can also exist that points from the Au to the Cu nucleus with a strength comparable to the magnetic field~\cite{Hirono:2012rt,Deng:2014uja,Voronyuk:2014rna}. The electric field can also lead to anomalous transport (i.e., the CESE~\cite{Huang:2013iia}; see also the derivation in holographic models~\cite{Pu:2014cwa,Bu:2018vcp} and discussion in Weyl semimetal~\cite{Zyuzin:2018etr}). The CESE is not directly related to the chiral anomaly and its appearance requires both P and C violations. The CESE represents an axial current along the direction of the electric field. Its expression for two flavor QCD up to leading-log accuracy is given by~\cite{Jiang:2014ura}
\begin{eqnarray}
{\vec J}_A\approx 14.5163\Tr (Q_e Q_A)\frac{\m_V\m_A}{T^2} \frac{eT}{g^4\ln (1/g)}\vec E,
\end{eqnarray}
where $Q_e$ and $Q_A$ are the charge and axial matrices in flavor space, and $g$ is the strong coupling constant. Of course, in addition to the CESE, the electric field induces the Ohm current $\vec J_V=\s \vec E$, where $\s$ is the electric conductivity, which, for QGP, is actually very high, meaning that the QGP is a good conducting matter~\cite{Ding:2016hua}.

Interesting collective modes emerge from the coupled evolution of the axial and vector charges through CME and CSE, vector CVE and axial CVE, or CESE and the usual Ohm's law. For example, the continuity equations for vector and axial charges can be written in terms of RH and LH charges:
\begin{eqnarray}
\label{continuation}
\pt_t J^0_{R/L}+{\vec\nabla}\cdot\bJ_{R/L} &=&0.
\end{eqnarray}
Substituting the CME and CSE expressions and considering small fluctuations in $J^0_{R,L}$ and $\m_{R,L}$, we obtain
\begin{eqnarray}
\label{waveequ}
\pt_t\d J^0_R+\frac{e^2}{4\p^2\c}\bB\cdot\vec\nabla\d J^0_R &=&0,\\
\label{waveequ2}
\pt_t\d J^0_L-\frac{e^2}{4\p^2\c}\bB\cdot\vec\nabla\d J^0_L &=&0,
\end{eqnarray}
where $\c=\pt J^0_R/\pt\m_R\approx\pt J^0_L/\pt\m_L$ is the number susceptibility, and we keep only linear terms in fluctuations. These two equations express two collective, gapless, wave modes, which are called chiral magnetic waves (CMWs)~\cite{Kharzeev:2010gd}. Similarly, if we consider the CESE and Ohm's law, we can find new collective modes, chiral electric waves, and axial or vector density waves~\cite{Huang:2013iia}. If we consider the vector and axial CVEs, we can find chiral vortical waves (CVWs)~\cite{Jiang:2015cva} described by $\pt_t \d J^0_{R/L}\pm v_{\rm CVW}\pt_z\d J^0_{R/L}=0$ with $v_{\rm CVW}=\m_{V0}\o/(2\p^2\c)$ being the propagating velocity of the CVWs. Note that, different from the CMWs, the occurrence of CVWs requires background vector density (characterized by $\m_{V0}$). Finally, we summarize the ACTs (and the usual Ohm's law) in Table \ref{tab-1}.

\begin{table*}
  \renewcommand{\arraystretch}{1.6}
\caption{Anomalous chiral transports}
\label{tab-1}
\begin{tabular*}{\hsize} {@{\extracolsep{\fill}}cccc}
\hline
  & $e\vec E$ & $e\vec B$ & $\vec \o$  \\\hline
$\displaystyle\vec J_V$ & $\displaystyle\s$ & $\displaystyle\frac{\m_A}{2\p^2}$ & $\displaystyle\frac{\m_V\m_A}{\p^2}$ \\[0.1cm]\hline
$\displaystyle\vec J_A$ & $\displaystyle\propto\frac{\m_V\m_A}{T^2}\s$ & $\displaystyle\frac{\m_V}{2\p^2}$ & $\displaystyle\frac{T^2}{6}+\frac{\m_V^2+\m_A^2}{2\p^2}$ \\[0.1cm]\hline
\;\;Collective mode\;\; & \;\;chiral electric wave\;\; & \;\;chiral magnetic wave\;\; & \;\;chiral vortical wave\;\; \\[0.1cm]\hline
\end{tabular*}
\end{table*}

\section{ACTs in heavy-ion collisions}
\label{sec-hic}
ACTs have attracted considerable attention in many subfields of physics, including nuclear physics, particle physics, astrophysics, condensed matter physics, atomic physics, and quantum optics. For heavy-ion collisions, in particular, ACTs provide a valuable means to detect the possible P and CP violations of QCD at high temperatures. It is a well-known experimental fact that the strong interaction is P and CP invariant in vacuum, although QCD itself permits the existence of P and CP violating $\h$ term. This lacks a natural explanation and is one of the main puzzles in contemporary physics. It has been proposed that in a high-temperature environment produced by heavy-ion collisions, metastable domains leading to P and CP violations could be produced through, for example, sphaleron-induced transition between gauge field vacua of different topological winding numbers~\cite{Kharzeev:1998kz,Kharzeev:2001ev,Kharzeev:2004ey}. In these domains, the interaction between gluons and quarks (through triangle anomaly) can induce chirality imbalance in quarks, which can be characterized by the parameter $\m_A$. Thus, the EM fields or vorticity exerting to these domains cause the CME, CVE, and CESE. Therefore, the detection of ACTs is highly demanded in heavy-ion collisions.

\subsection {Experimental search of CME}\label{sec:corrcme}
Because the magnetic field is roughly perpendicular to the reaction plane, the CME would drive a current that finally causes a charge separation with respect to the reaction plane. However, the production of $\m_A$ has strong spatial fluctuation (among the metastable P-violating domains) and event-by-event fluctuation such that the event-averaged CME-induced charge separation vanishes. What can be observed is the fluctuation of the charge separation. This can be done by designing appropriate hadronic observables. One commonly used observable is the $\gamma$ correlation introduced by Voloshin~\cite{Voloshin:2004vk}:
\begin{eqnarray}
\label{gamma}
\g_{\a\b}\equiv\lan\cos(\f_\a+\f_\b-2\J_\rp)\ran,
\end{eqnarray}
where $\a,\b=\pm$ denote the charge signs, $\f_\a$ and $\f_\b$ are the corresponding azimuthal angles, $\J_\rp$ is the reaction plane angle, and $\lan\cdots\ran$ is the event average. It is easy to see that a charge separation with respect to the reaction plane results in positive $\g_{+-}$ and $\g_{-+}$ (denoted as $\g_{\rm OS}$) and negative $\g_{++}$ and $\g_{--}$ (denoted as $\g_{\rm SS}$). In real experiments, one additional reference hadron (of arbitrary charge) is required to determine $\J_\rp$. Therefore, \eq{gamma} is practically a three-particle correlation.

\begin{figure}[!htb]
\begin{center}
\includegraphics[width=7cm]{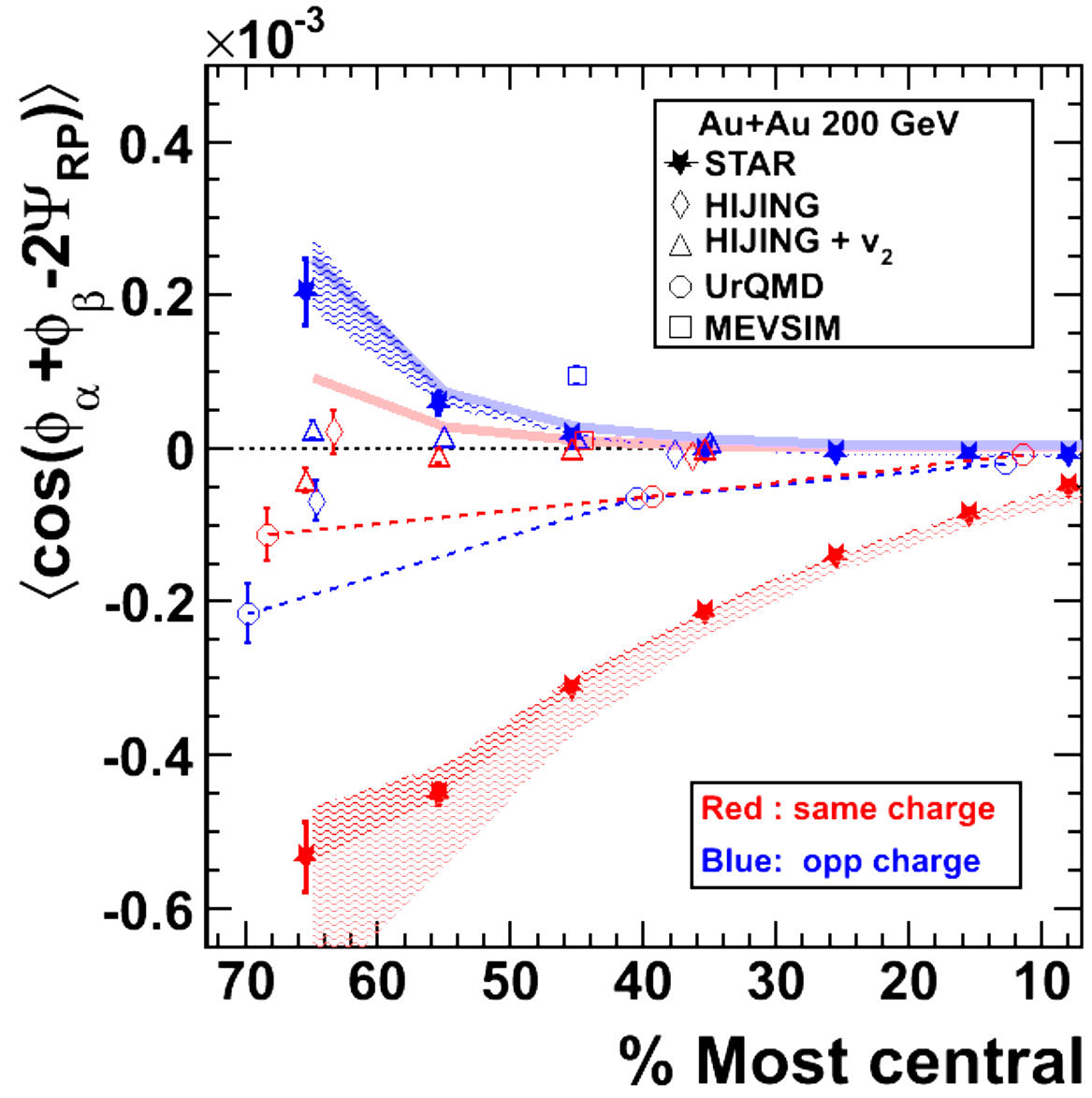}
\caption{Correlation $\g_{\a\b}$ measured by the STAR Collaboration at RHIC. The figure is from Ref.~\cite{Abelev:2009ac}.}
\label{expcme}
\end{center}
\end{figure}

The correlation $\g_{\a\b}$ was first measured by the STAR Collaboration at RHIC for Au + Au collisions at $\sqrt{s}=200$ GeV~\cite{Abelev:2009ac,Abelev:2009ad}; see \fig{expcme}. The same quantity was also measured by: 1) ALICE Collaboration at LHC for Pb + Pb collisions at $\sqrt{s}=2.76$ TeV~\cite{Abelev:2012pa,Acharya:2017fau}, 2) CMS Collaboration at LHC for Pb + Pb collisions at $\sqrt{s}=5.02$ TeV~\cite{Khachatryan:2016got,Sirunyan:2017quh}, and 3) STAR Collaboration for Au + Au collisions at different beam energies down to $\sqrt{s}=19.6$ GeV~\cite{Adamczyk:2014mzf}. For mid-central collisions, these measurements show positive $\g_{\rm OS}$ and negative $\g_{\rm SS}$ with features consistent with the expectation of CME. However, non-CME background effects exist in the $\g$ correlation, noticeably, the transverse momentum conservation (TMC) and local charge conservation (LCC). Before a convincing means of subtracting these backgrounds can be obtained, we cannot claim an observation of the CME. The TMC induces a back-to-back correlation to $\g_{\a\b}$~\cite{Pratt:2010zn,Bzdak:2010fd}, which can be subtracted by making a difference $\D\g\equiv\g_{\rm OS}-\g_{\rm SS}$, as the TMC is charge blind. The LCC is more difficult to subtract~\cite{Schlichting:2010qia,Wang:2009kd}, which gives a finite contribution to $\D\g$, namely, $\D\g^{\rm LCC}\propto M v_2/N$,
where $M$ is the number of hadrons in a local neutral cell, $N$ is the multiplicity, and $v_2$ is the elliptic flow.

The main challenge remaining with the experiments is to disentangle the elliptic-flow-driven background effects and the magnetic-field-driven CME signal. One important experimental progress is the measurement of the $\g$ correlation in small systems such as p(d) + A collisions. In p(d) + A collisions, although the magnetic field could be large, its orientation is not correlated to the participant plane (or $v_2$ plane). Thus, the magnetic field is not expected to drive a strong $\g$ correlation measured with respect to the $v_2$ plane. Therefore, the p(d) + A collisions can serve as a baseline for the background contributions. The recent results from CMS ~\cite{Khachatryan:2016got,Sirunyan:2017quh} and STAR \cite{STAR:2019xzd} Collaborations showed that the $\g$ correlation in p(d) + A collisions is comparable to or even larger than that in A + A collisions at the same energy and multiplicity. This suggests that the $\g$ correlation contains a large portion of background contribution for peripheral A + A collisions; see additional discussions in Refs. ~\cite{Khachatryan:2016got,Sirunyan:2017quh,STAR:2019xzd,Li:2020dwr}.

\begin{figure}[!htb]
\begin{center}
\includegraphics[width=7cm]{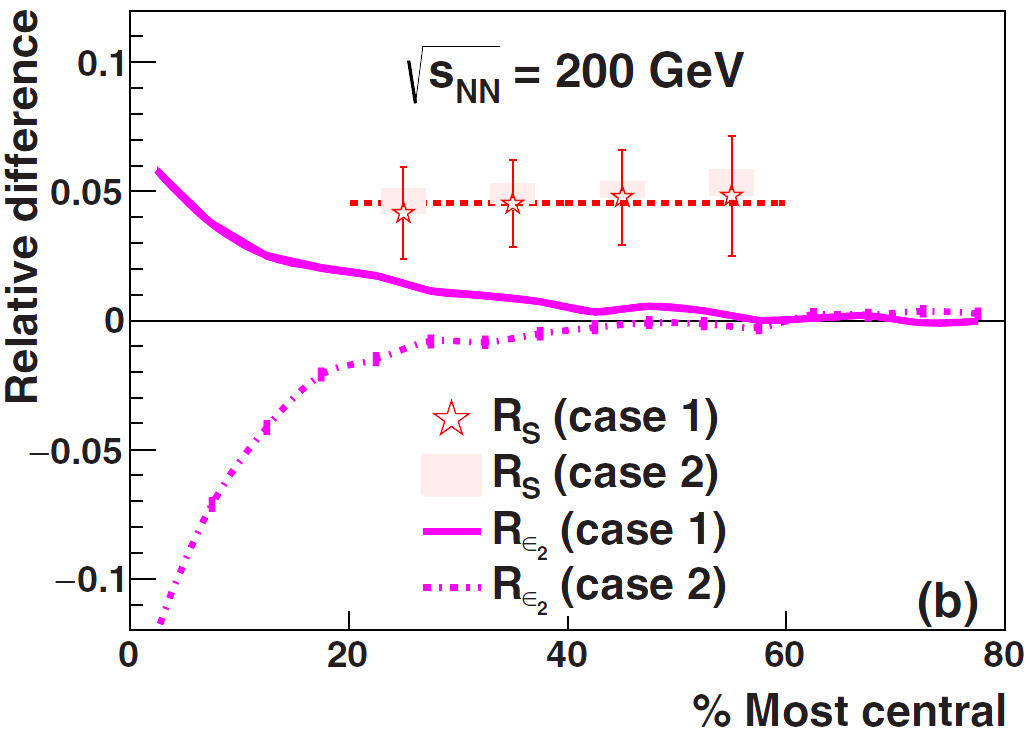}
\caption{Relative difference in eccentricity $R_{\e_2}$ and corrected $\g$ correlation $R_S$ for background level ${\rm bg}=2/3$ in isobar collisions with $400$ million events for each collision type. The figure is from Ref.~\cite{Deng:2016knn}.}
\label{isobars}
\end{center}
\end{figure}

Another important experimental progress, namely, the isobar collision was made in 2018 at RHIC. In this experimental program, two sets of collisions are operated, one for $^{96}_{44}$Ru + $^{96}_{44}$Ru and the other for $^{96}_{40}$Zr + $^{96}_{40}$Zr~\cite{Voloshin:2010ut,Deng:2016knn,Huang:2017azw,Huang:2017azw,Xu:2017zcn,Li:2018oec,Sun:2018idn,Magdy:2018lwk,Shi:2019wzi}. It is expected that these two collisions with the same beam energy and same centrality will produce roughly equal elliptic flow but a $10\%$ difference in magnetic fields. If $\D\g$ contains a contribution from CME, we should see a difference in $\D\g$ between Ru + Ru and Zr + Zr collisions. To quantify the sensitivity of the isobar collisions, let us define the relative difference of the eccentricity $R_{\e_2}=2(\e_2^{{\rm Ru}+{\rm Ru}}-\e_2^{{\rm Zr}+{\rm Zr}})/(\e_2^{{\rm Ru}+{\rm Ru}}+\e_2^{{\rm Zr}+{\rm Zr}})$ (note that $v_2$ is usually proportional to $\e_2$). Similarly, we can define $R_{B_{\rm sq}}$ to quantify the relative difference in the projected magnetic field squared $B_{\rm sq}\equiv\lan(eB/m_\p^2)^2\cos[2(\J_{\rm B}-\J_{\rm RP})]\ran$ (with $\J_{\rm B}$ being the azimuthal angle of the magnetic field)~\cite{Bloczynski:2012en,Bloczynski:2013mca} and $R_S$ to quantify the relative difference in the corrected $\g$ correlation $S=N_{\rm part}\D\g$ (where $N_{\rm part}\propto N$ is the participant number used to compensate for the dilution effect). Because $R_{\e_2}$, $R_{B_{\rm sq}}$, and $R_S$ are small, we can take a linear approximation to link them, that is, $R_S=(1-{\rm bg})R_{B_{\rm sq}}+{\rm bg} R_{\e_2}$. The quantities $R_{\e_2}$ and $R_{B_{\rm sq}}$ can be easily obtained from theoretical simulation. We can then obtain $R_S$ as a function of the background level $\rm bg$ through this relation. In \fig{isobars}, we show the numerical results for $R_{\e_2}$ and $R_S$ for ${\rm bg}=2/3$ with $400$ million events for each collision type~\cite{Deng:2016knn}. In this situation, the significance level of the discovery of the CME signal reaches $5\s$ for centrality region $20-60\%$. In the 2018 experiment, the total number of collision events was $6.3$ billion~\cite{Adam:2019fbq} and a $5\sigma$ significance level of the discovery of CME could be reached even for $\rm bg\approx 88\%$ or a $3\s$ significance level for $\rm bg\approx 93\%$ in centrality region $20-60\%$~\footnote{We thank G. Wang for discussion on this topic.}. Currently, the STAR Collaboration is conducting a blind analysis of the isobar data, and we are really looking forward to their results.

Recently, other methods have been proposed for the purpose of disentangling the CME signal and the backgrounds. They include the pair invariant mass dependence of the $\g$ correlation~\cite{Zhao:2017nfq,Zhao:2018blc}, a comparative measurement of the $\g$ correlation with respect to reaction and participant planes~\cite{Xu:2017qfs,Zhao:2018blc}, the signed balance functions~\cite{Tang:2019pbl}, and the charge-sensitive in-event correlations~\cite{Magdy:2017yje}. A detailed discussion can be found in the cited studies.

\subsection {Experimental search of other ACTs}\label{sec:otheract}
The chiral magnetic wave can transport both the vector and axial charges and can lead to an electric quadrupole in the QGP with more positive charges on the tips of the fireball and more negative charges in the equator of the fireball~\cite{Gorbar:2011ya,Burnier:2011bf}. Therefore, hydrodynamic expansion of the fireball drives a larger $v_2$ for negative charges (for example, $\p^-$) than the positive charges (for example, $\p^+$)~\cite{Burnier:2011bf,Burnier:2012ae,Taghavi:2013ena,Yee:2013cya,Hirono:2014oda}.
The difference $\D v_2=v_2(\p^{-})-v_2(\p^+)$ is proportional to the net charge asymmetry $A_{\rm ch}=(N_+-N_-)/(N_++N_-)$; this is because the CSE is proportional to $\m_V$. This charge dependence of $v_2$ was measured by the STAR Collaboration~\cite{Adamczyk:2015eqo} at RHIC and by ALICE Collaboration~\cite{Adam:2015vje} and CMS Collaboration~\cite{Sirunyan:2017tax} at LHC. The data show an elliptic-flow difference $\D v_2$ linear in $A_{\rm ch}$ with a positive slope whose centrality dependence is consistent with the expectation of the CMW. However, we should emphasize that, similar to the measurement of the $\g$ correlation, non-CMW background effects exist, that contribute to $\D v_2$~\cite{Deng:2012pc,Stephanov:2013tga,Dunlop:2011cf,Xu:2012gf,Song:2012cd,Bzdak:2013yla,Hatta:2015hca,Xu:2019pgj}. A conclusive claim about the experimental results for the CMW search can be made only after we can successfully subtract the background effects, which we are unable to do now.

\begin{figure}[!htb]
\begin{center}
\includegraphics[width=7cm]{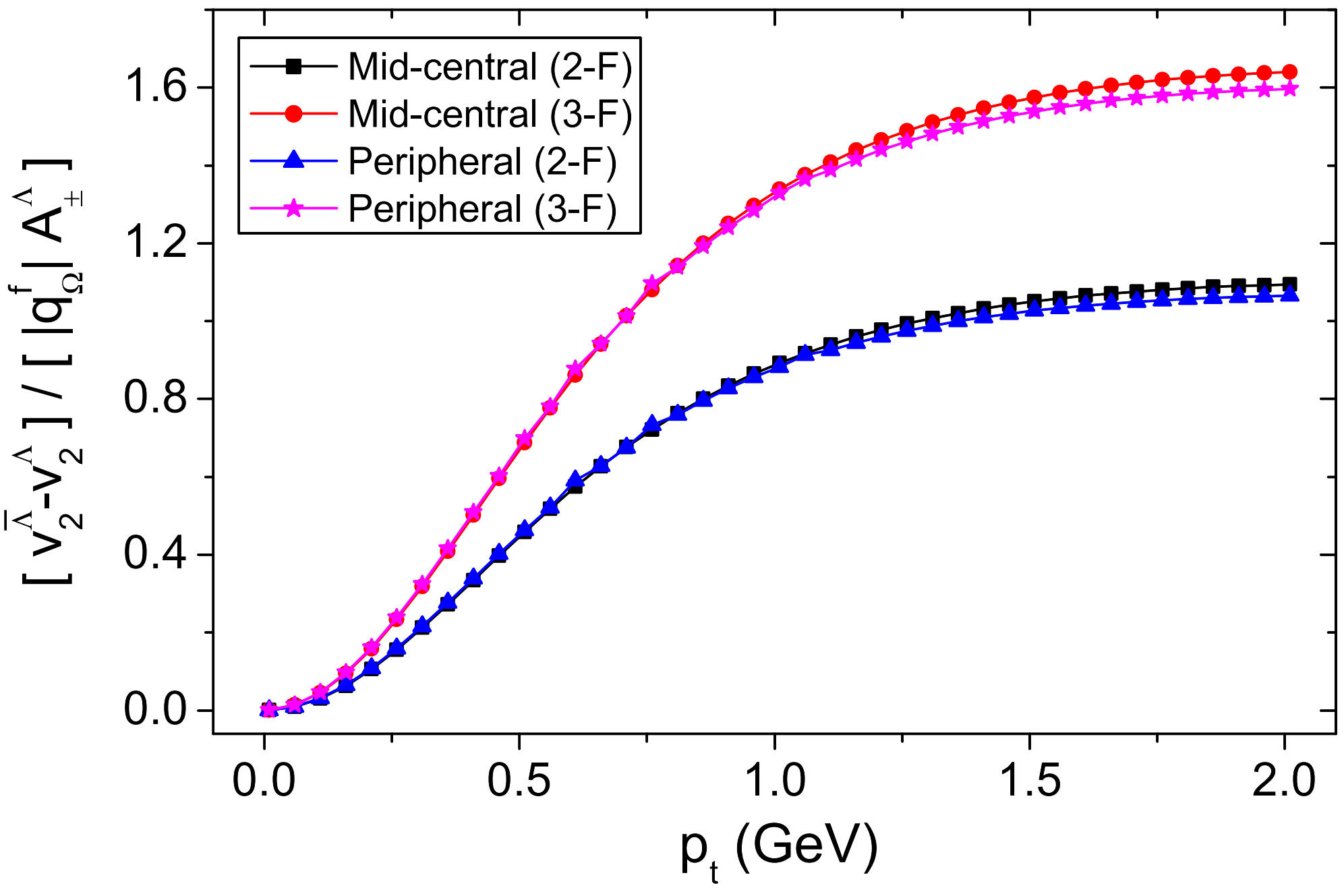}
\caption{Splitting of $v_2$ between $\L$ and $\bar{\L}$ induced by the chiral vortical wave. The figure is from Ref.~\cite{Jiang:2015cva}.}
\label{figcvw}
\end{center}
\end{figure}

In heavy-ion collisions, the transverse space-averaged vorticity at the mid-rapidity region is roughly perpendicular to the reaction plane. Therefore, similar to the CME case, the vector CVE induces a baryon number separation with respect to the reaction plane. We can use a correlation similar to the $\g$ correlation for CME to detect the vector-CVE-induced baryon number separation (i.e., $\w_{\a\b}=\lan\cos(\phi_\a+\phi_\b-2\Psi_{\rm RP})\ran$, where $\a,\b=\pm$ denote baryons or anti-baryons and $\phi_{\a,\b}$ is the corresponding azimuthal angle). However, similar to what occurs with the CME search, it would be challenging to subtract the possible background contributions as with the transverse momentum conservation and local baryon number conservation in the $\w$ correlation. The implication of the CVW in heavy-ion collisions is that it could induce a baryon quadrupole in the QGP in such a manner that more baryons and anti-baryons are distributed on the tips and in the equator of the fireball, respectively. After the collective expansion of the fireball, the baryons (for example, $\L$) would have smaller $v_2$ than the anti-baryons ($\bar{\L}$) with the difference being proportional to the net baryon asymmetry $A_\pm^{\L}=(N_\L-N_{\bar\L})/(N_\L+N_{\bar\L})$; see \fig{figcvw} for a theoretical simulation of $v_2(\bar{\L})-v_2(\L)$ versus $p_t$~\cite{Jiang:2015cva}. As the produced $\L$ and $\bar{\L}$ are considerably rarer than $\p^\pm$, the detection of this difference is statistically more challenging than $v_2(\p^-)-v_2(\p^+)$. We expect that phase II of the RHIC beam energy scan program would provide a new possibility for the search of CVE and CVW~\cite{Bzdak:2019pkr}.

\begin{figure}[!htb]
\begin{center}
\includegraphics[width=7cm]{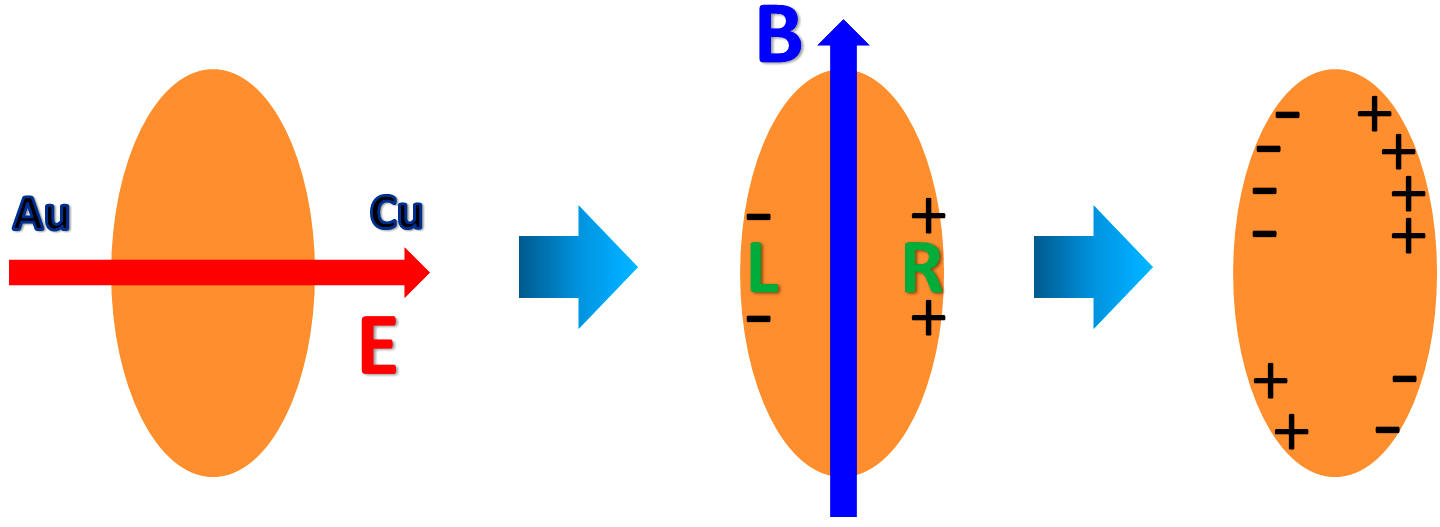}
\caption{Quadrupolar pattern of charge distribution induced by the CESE and CME in Cu + Au collisions. The figure is from Ref.~\cite{Huang:2013iia}.}
\label{figcese}
\end{center}
\end{figure}

The non-central Cu + Au collisions may be used to test the CESE, as they generate a persistent electric field orientating from the Au to Cu nuclei~\cite{Deng:2014uja}. As illustrated in \fig{figcese}, the CESE induces an axial charge separation along the impact parameter direction (e.g., RH and LH chiralities on the near-Cu and near-Au sides, respectively), the CME in turn induces a charge separating pattern as shown in the last step (which is superposed by an Ohm-current-induced in-plane charge separation). A possible observable for this special quadrupolar pattern of charge distribution can be the charge dependence of the event planes, namely, a finite $\D\Psi=\lan|\Psi_2^+-\Psi_2^-|\ran$ increasing with the centrality, where $\Psi_2^\pm$ is the event plane reconstructed from positively/negatively charged hadrons~\cite{Ma:2015isa}. Another possible observable is the $\z$ correlation~\cite{Ma:2015isa}, $\z_{\a\b}=\lan\cos[2(\phi_\a+\phi_\b-2\Psi_{\rm RP})]\ran$. However, we should note that as the CESE is proportional to $\m_V\m_A/T^2$, which is small for typical heavy-ion collisions, the test of CESE requires numerous collision events.

\section{Spin polarization in heavy-ion collisions}
\label{sec:sp}
A remarkable effect of vorticity is that it could polarize the spin of the constituent particles~\cite{Liang:2004ph,Voloshin:2004ha,Gao:2007bc,Huang:2011ru}. This is simply due to the quantum mechanical spin--orbit coupling. The motion of the fluid cell with finite vorticity generates an orbital angular momentum that can be transferred to the spin degree of freedom of the particles that constitute the fluid. If the system attains thermal equilibrium, we can use statistical mechanics to estimate the spin polarization. The density operator is $\hat{\r}=Z^{-1}\exp\ls-\b( \hat{H}- \hat{\bm S}\cdot\bm \o)\rs$, where $\bm \o$ is the nonrelativistic vorticity, $\hat{H}$ the spin-unpolarized Hamiltonian, $\hat{\bm S}$ the spin operator (with the orbital-angular-momentum part being absorbed in the $\b\hat{H}$ term), and $Z$ the partition function. The spin polarization is given by $\bm P=\Tr[\hat{\bm S}\hat{\r}]/s$, where $s$ is the spin quantum number. For fermions of spin $1/2$, we have $\hat{\bm S}=\bm \s/2$ with $\bm\s$ the Pauli matrices, and thus, $\bm P=\bm\o/(2T)+o(\o/T)$. The more rigorous derivation shows that, for spin-1/2 fermions, the spin four-vector is given as~\cite{Becattini:2013fla,Fang:2016vpj,Florkowski:2017dyn,Liu:2020flb}
\begin{eqnarray}
\label{spin}
S^\m(x,p)=-\frac{1}{8m}(1-n_F)\e^{\m\n\r\s}p_\n\varpi_{\r\s}(x)+O(\varpi^2),
\end{eqnarray}
where $n_F(p_0)$ with $p_0=\sqrt{\bm p^2+m^2}$ is the Fermi--Dirac distribution function and $\varpi_{\r\s}(x)$ is the thermal vorticity. For $\L$ and $\bar\L$ hyperons, $s=1/2$, and we have approximately $1-n_F\approx 1$, as they are heavy. In the rest frame of the particle, $S^{*\m}=(0,\bm S^*)$, where $\bm S^*$ can be obtained by using Lorentz transformation,
\begin{eqnarray}
\bm S^{*}=\bm S-\frac{\bm p\cdot\bm S}{p_0(p_0+m)}\bm p.
\end{eqnarray}
Thus, we obtain the polarization vector in the rest frame of the particle as
\begin{eqnarray}
\label{polar}
\bm P^*=\frac{\bm S^*}{s}.
\end{eqnarray}
In the following, without confusion, we simply use $\bm P$ to denote the polarization vector in the rest frame.
\begin{figure}[!htb]
\begin{center}
\includegraphics[width=7cm]{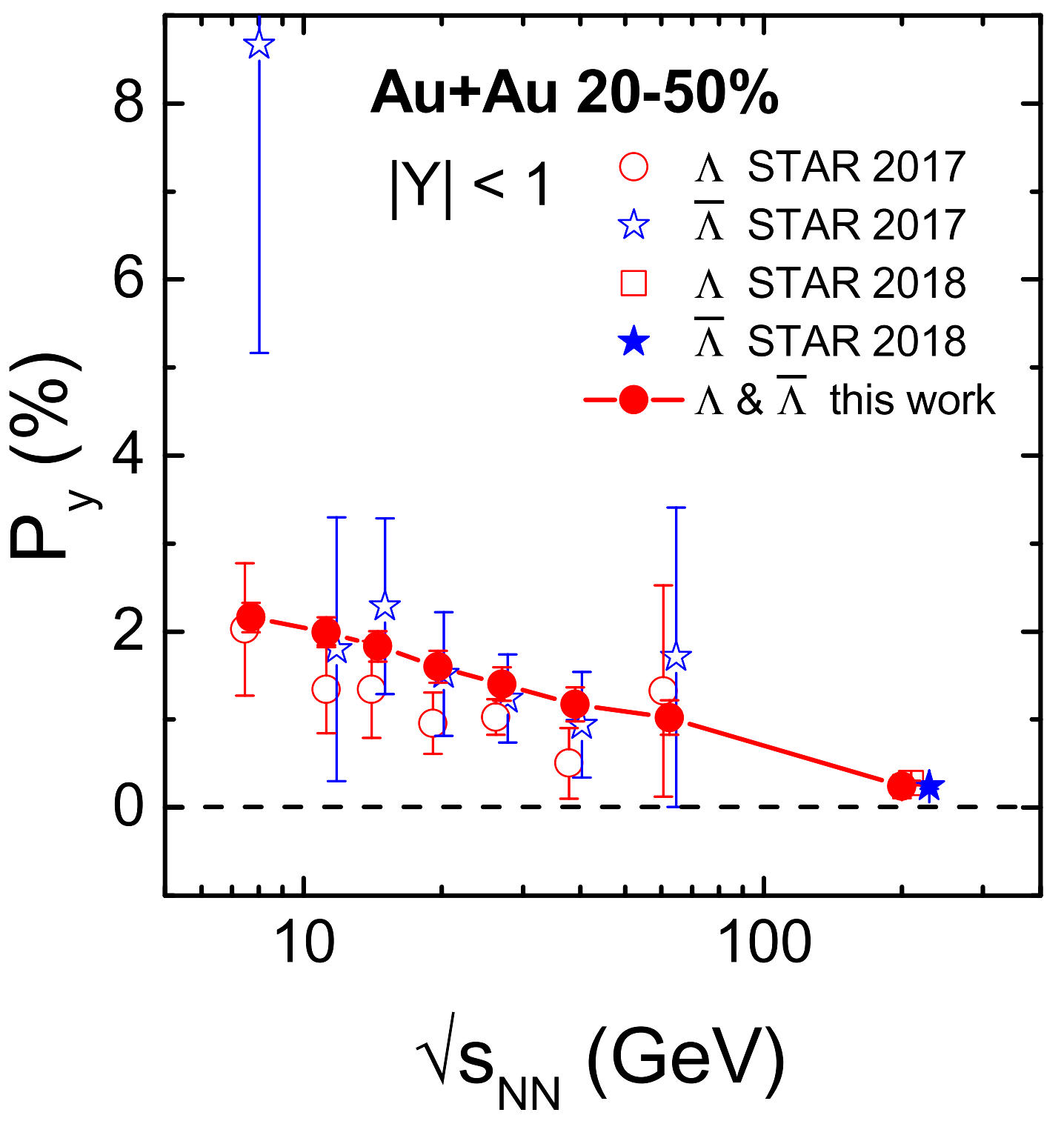}
\caption{Global $\L$ polarization at mid-rapidity in Au + Au collisions. The figure is from Ref.~\cite{Wei:2018zfb}.}
\label{figpy}
\end{center}
\end{figure}

Before discussing the experimental measurements and numerical computations, let us explain the relation and distinction between the spin polarization of hyperons and ACTs in heavy-ion collisions. The ACTs are closely related to the chiral anomaly of QCD and/or QED, which is critical in modern physics. Detecting the ACTs also provides strong evidence for chiral symmetry restoration in the hot QGP. However, the underlying mechanism of the spin polarization is not related to the chiral anomaly but to quantum mechanical spin-orbit coupling. Importantly, spin polarization measurements provide a new probe for the QGP, that is, the spin probe, which is complementary to the usual probes using, for example, the charges.
The ACTs and spin polarization of hyperons are also closely related to each other. First, they all represent the responses of the hot medium to the external vortical or EM field. In fact, as we will see in the following, the spin polarizations of $\L$ and $\bar{\L}$ are not identical, which probably reflects the response to the magnetic field. Second, as we discussed in \sect{sec:anom}, using QED as an example, the chiral anomaly is also understood as a type of spin polarization; the spin is fully polarized in the LLL, which is responsible for the chiral anomaly. Therefore, the ACTs and spin polarization of hyperons provide different angles to observe how the spin degree of freedom in the medium responds to the vortical and EM fields. Thus, we discuss them together in this article.

Substituting the theoretically calculated thermal vorticity shown in \fig{the-omg} into \eqs{spin}{polar}, we can obtain the $y$ component of the spin polarization. This reflects the global angular momentum of the collision system and is called the global spin polarization; see \fig{figpy} for global spin polarization of $\L$ and $\bar{\L}$ (in short, ``$\L$ polarization")~\cite{Wei:2018zfb}. In addition, the experimental data from the STAR Collaboration are also shown~\cite{STAR:2017ckg,Adam:2018ivw}. We find that the theoretical results fit the data very well. We note that similar calculations were performed by using either transport or hydrodynamic models, and good matches with the experimental data were seen in all those calculations~\cite{Karpenko:2016jyx,Xie:2017upb,Li:2017slc,Sun:2017xhx,Shi:2017wpk,Csernai:2018yok,Wei:2018zfb,Xie:2019npz,Ivanov:2019wzg}. From \fig{figpy}, we can see two special features of the global $\L$ polarization. One is that the global $\L$ polarization (as well as the vorticity) is smaller for larger $\sqrt{s}$. This contradicts our intuition because the total angular momentum of the system should be greater for larger $\sqrt{s}$. This is a relativistic effect: with increasing collision energy, the created hot matter at mid-rapidity behaves increasingly boost-invariant along the beam direction, thus supporting gradually less vorticity at mid-rapidity~\cite{Deng:2016gyh,Jiang:2016woz}. However, for a very-low-energy collision, the system may be nonrelativistic and the initial vorticity, which well reflects the angular momentum retained in the mid-rapidity region, would increase with $\sqrt{s}$~\cite{Deng:2020ygd}. The other feature is as follows: despite a big error bar, the experimental data show that $\L$ spin polarization is less than $\bar{\L}$ spin polarization. Some possibilities for this difference have been recently discussed~\cite{Han:2017hdi,Guo:2019joy,Guo:2019mgh,Csernai:2018yok}. For example, as the Zeeman coupling between the magnetic field and spin depends on the magnetic moment of the particle, $\bar{\L}$, which has a positive magnetic moment, is more easily polarized than $\L$, which has a negative magnetic moment.

\begin{figure}[!htb]
\begin{center}
\includegraphics[width=7cm]{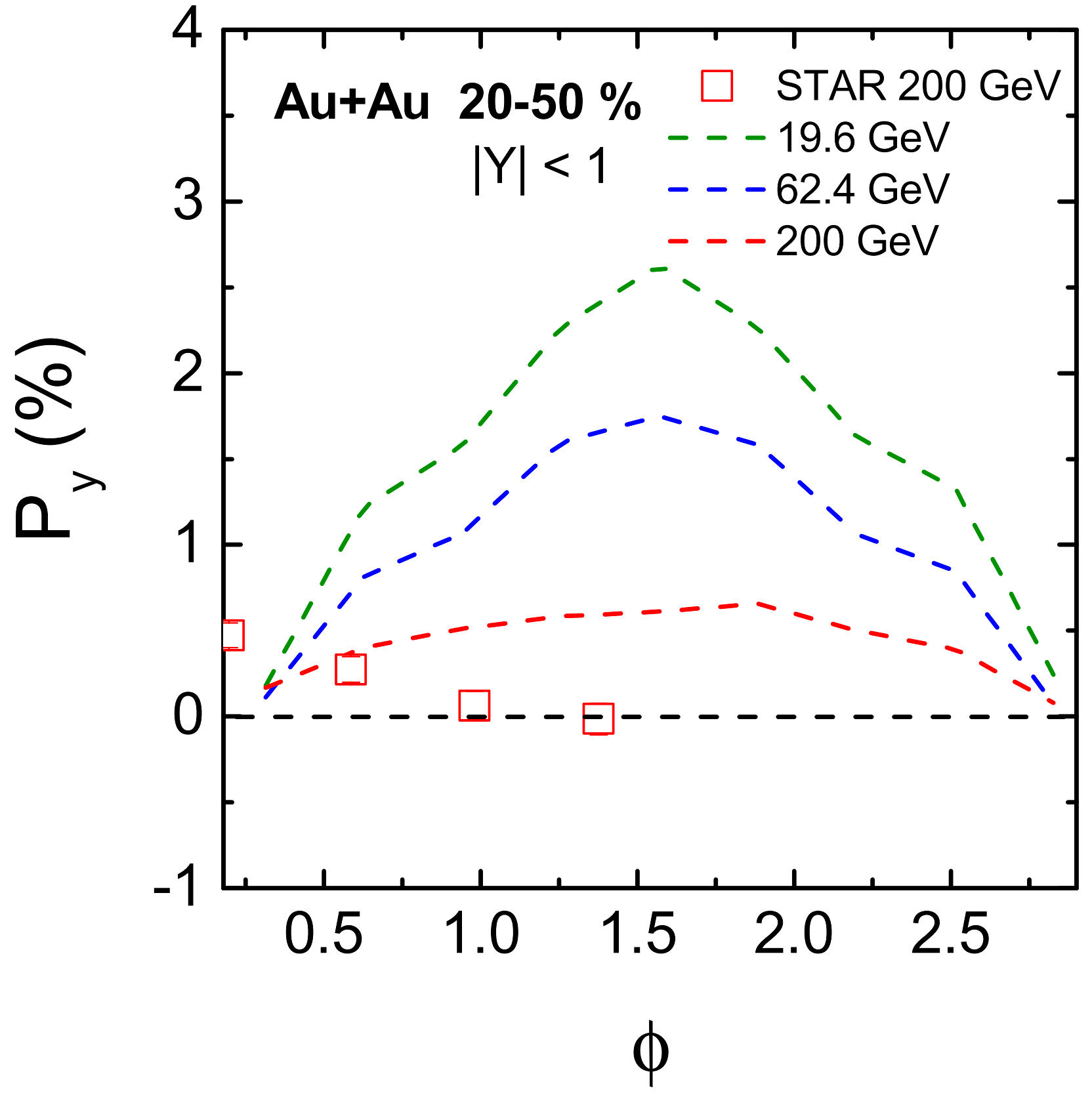}
\caption{$\L$ polarization along the $y$ direction as a function of the azimuthal angle at mid-rapidity. The figure is from Ref.~\cite{Wei:2018zfb}.}
\label{figpy-phi}
\end{center}
\end{figure}

\begin{figure}[!htb]
\begin{center}
\includegraphics[width=7cm]{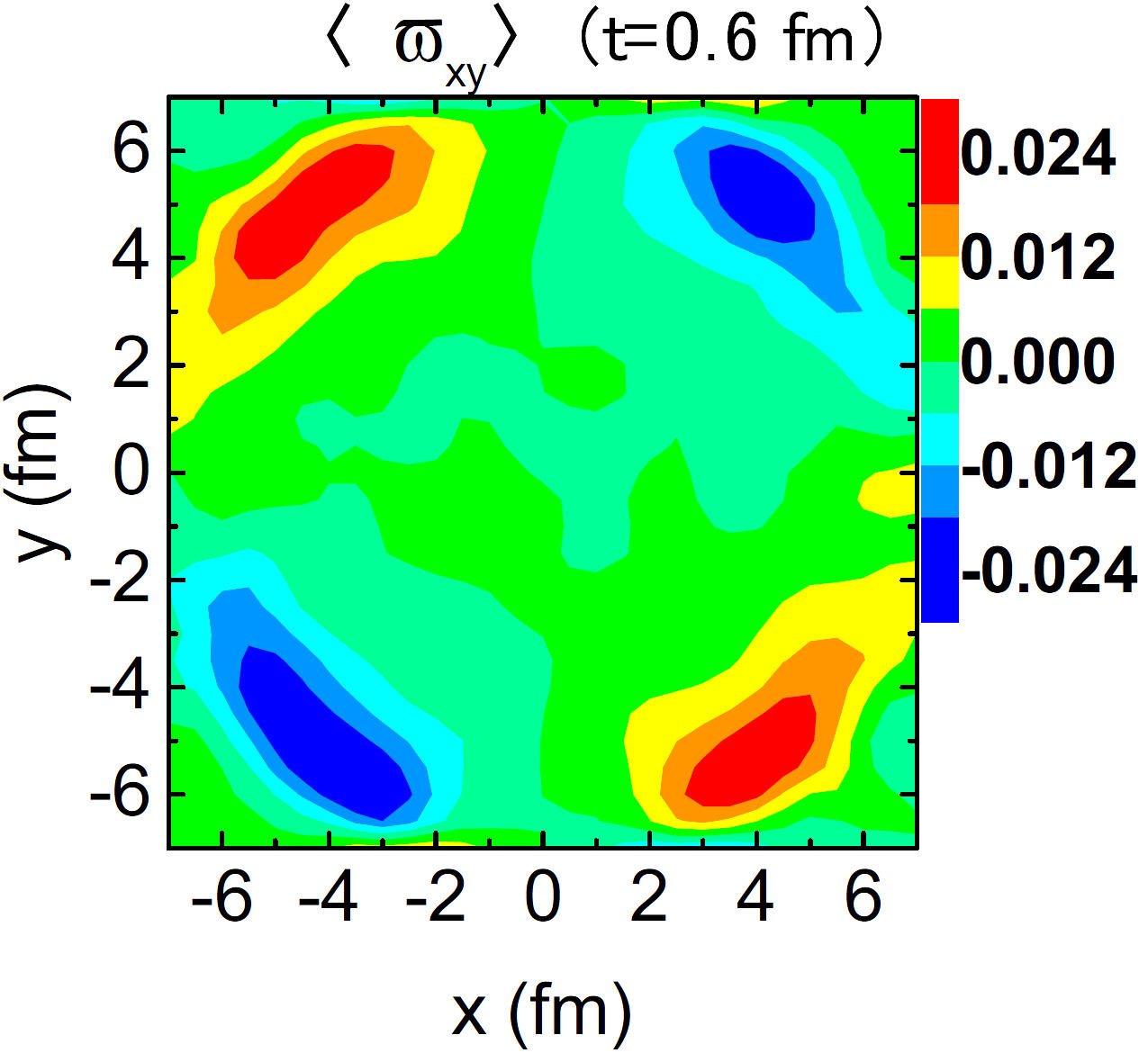}
\caption{Longitudinal thermal vorticity $\varpi_{xy}$ in the $x$-$y$ plane ($z=0$) in a non-central heavy-ion collision. The figure is from Ref.~\cite{Wei:2018zfb}.}
\label{figlong}
\end{center}
\end{figure}

Recently, the STAR Collaboration published their measurements of differential spin polarization, namely, the dependence of $\L$ polarization on the kinematic variables such as the azimuthal angle and transverse momentum~\cite{Adam:2018ivw,Adam:2019srw}. In describing the differential spin polarization, the theoretical calculations thus far have been unsatisfactory. In particular, the calculations based on hydrodynamic and transport models show that $P_y(\phi)$ ($\phi$: azimuthal angle) at mid-rapidity increases when $\phi$ grows from $0$ to $\p/2$. However, the experimental data show the opposite; see \fig{figpy-phi}~\cite{Wei:2018zfb}. Similarly, for noncentral collisions, a nonzero longitudinal $\L$ polarization $P_z(\phi)$ is observed in experiments (where this polarization vanishes when integrated over all the angles $\phi$), indicating a $\phi$ dependence that is also qualitatively opposite to the theoretical calculations of thermal vorticity~\cite{Becattini:2017gcx,Xia:2018tes,Wei:2018zfb}; see \fig{figlong}. Expressed in formula as
\begin{eqnarray}
\label{spin-har}
\begin{split}
\frac{dP_{y,z}}{d\phi}=P_{y,z}&+2 f_{2y,z}\sin[2(\phi-\Psi_{\rm RP})]\\
&+2 g_{2y,z}\cos[2(\phi-\Psi_{\rm RP})]+\cdots,
\end{split}
\end{eqnarray}
the second-order harmonic coefficient $f_{2z}$ (and $g_{2y}$) has the opposite sign in current theoretical calculations and in experimental data (i.e., $f_{2z}^{\rm ther}<0, g_{2y}^{\rm ther}<0$ while $f_{2z}^{\rm exp}>0, g_{2y}^{\rm exp}>0$). This is a huge puzzle. We call it the ``spin sign problem.¡¯¡¯ To resolve the spin sign problem, some important issues should be carefully re-examined. (1) Approximately $80\%$ of the measured $\L$ and $\bar{\L}$ are from decays of other higher-lying hadrons. During these decays, it is possible (e.g., in $\S^0\ra\L+\g$) that the spin-polarization direction of the daughter $\L$ is flipped as compared with the parent particle. Recent studies have shown that these decay contributions, despite suppressing $\sim 10\%$ of the primary $\L$ polarization, are insufficient to resolve the spin sign problem~\cite{Xia:2019fjf,Becattini:2019ntv}. (2) Possible initial local spin polarization or an initial flow profile that can lead to finite local vorticity have not been encoded in hydrodynamic and transport models. It is a crucial future task to perform a numerical test of these possible initial conditions. (3) The formula (\ref{spin}) is derived based on the assumption that both momentum and spin degrees of freedom are at global equilibrium. This is a strong assumption that may not conform with a realistic case in heavy-ion collisions. Apart from global equilibrium, spin polarization is no longer enslaved to thermal vorticity and should be treated as an independent dynamical variable. Developing new theoretical framework that is beyond the global equilibrium assumption is very urgent. These frameworks, in both hydrodynamic and kinetic setups, have considerably progressed recently. We will discuss the hydrodynamic and kinetic frameworks with spin as a dynamical variable in the following sections. (4) Understanding the polarization dynamics is important. Recent studies include Refs.~\cite{Li:2018srq,Zhang:2019xya,Kapusta:2019sad,Ayala:2019iin,Ayala:2020ndx}. (5) Other issues that may influence the $\L$ polarization should also be explored (e.g., hadronic mean-fields~\cite{Csernai:2018yok}, chiral-anomaly-induced effects~\cite{Sun:2018bjl,Liu:2019krs}, other possible spin chemical potentials~\cite{Florkowski:2019voj,Wu:2019eyi,Xie:2019jun}, and the gluonic contribution). Testing complementary observables to measure the vorticity is also helpful (e.g., the $\phi$- and $K^{*0}$-spin alignment~\cite{Liang:2004xn}, the CVEs and CVW, and the recently proposed vorticity-dependent hadron yields~\cite{Taya:2020sej}).

To conclude this section, we explain how the special pattern of the thermal vorticity shown in \fig{figlong} emerges. Although in \sect{sec:fields} we discussed the fact that the global angular momentum of the collision system is the cause of vorticity, it is not the only cause. There are many other sources of vorticity. One important source is the inhomogeneous expansion of the fireball. Because in the non-central collisions, the fireball is almond shaped, the gradient of pressure would more strongly drive the fireball expanse along the reaction plane, and this is why we observe positive elliptic flow $v_2$. In this type of expansion, we can easily imagine that a vortical structure with four vortices in four quadrants of the $x$-$y$ plane ($z=0$) would appear. Of course, the temperature is also inhomogeneous and its gradient also contributes to the thermal vorticity, which together with the gradient of the velocity field gives the pattern shown in \fig{figlong}.

\section{Spin hydrodynamics}
\label{sec:sh}
Many attempts have been made to solve the spin sign problem. However, thus far, no satisfactory solution has been found. From a theoretical point of view, a key step forward would be to develop new theoretical frameworks to describe the spin polarization beyond the global equilibrium assumption. One promising framework is hydrodynamics, which very effectively describes the bulk evolution of the fireball in heavy-ion collisions, with the dynamical spin degree of freedom encoded. This type of framework is the spin hydrodynamics in which the spin polarization density (or equivalently the spin chemical potential) is treated on the same level as temperature $T$ and flow velocity $u^\m$~\cite{Florkowski:2017ruc,Montenegro:2018bcf,Florkowski:2018fap,Hattori:2019lfp,Bhadury:2020puc}.

In first-order spin hydrodynamics, the energy-momentum and spin current tensors are given by
\begin{eqnarray}
\label{spin-cons}
&T^{\m\n}=eu^\m u^\n-P\D^{\m\n}+\s_\w^{\m\n}+\s_\z^{\m\n}+2q^{[\m}u^{\n]}+\f^{\m\n},\nonumber\\
&\S^{\m,\a\b}=u^\m S^{\a\b},
\end{eqnarray}
in which we have chosen the Landau--Lifshitz frame. Here, $e$ is the energy density, $P$ is the pressure, $\s^{\m\n}_\w,\s^{\m\n}_\z$ are shear and bulk viscous tensors, $q^{\m}$ and $\f^{\m\n}=\f^{[\m\n]}$ are related to the spin degree of freedom and represent the strength of the torque on the temporal and spacial components of the spin current tensor. The constitutive relations are given as~\cite{Hattori:2019lfp}
\begin{eqnarray}
\label{spin-deco}
&\s_\w^{\m\n}=2\w\pt_\perp^{\lan\m}u^{\n\ran},\\
&\s_\z^{\m\n}=\z\h\D^{\m\n},\\
&q^\m=\l(D u^\m+\b\pt_\perp^\m T-4\O^{\m\n}u_\n),\\
\label{spin-deco2}
&\f^{\m\n}=2\g(\pt_\perp^{[\m}u^{\n]}+2\O^{\m\n}_\perp),
\end{eqnarray}
where $X^{[\a\b]}=(X^{\a\b}-X^{\b\a})/2$ is anti-symmetrization in indices $\a, \b$; $X^{\lan\a\b\ran}=(X^{\a\b}+X^{\b\a})/2-{X^\m}_\m\D^{\a\b}/3$ is traceless symmetrization in indices $\a, \b$; $\h=\pt_\m u^\m$ is the expansion rate; $\D_{\m\n}=g_{\m\n}-u_\m u_\n$ is the spatial projection; $D=u\cdot\pt$ is the co-moving time derivative; $\pt^\m_\perp=\D^{\m\n}\pt_\n$ is the spatial derivative; $\O^{\m\n}$ is called the spin chemical potential in which $\O^{\m\n}_\perp=\D_{\m\r}\D_{\n\s}\O^{\r\s}$. Here, $\w, \z, \l$, and $\g$ are the transport coefficients, which must be semi-positive. We call them the shear viscosity, bulk viscosity, boost heat conductivity, and rotational viscosity, respectively~\cite{Hattori:2019lfp}. A recent attempt at calculating the spin-related transport coefficient is given in Ref.~\cite{Li:2019qkf}. The hydrodynamic equations are as follows:
\begin{eqnarray}
\label{spin-hydro}
&\pt_\m T^{\m\n}=0,\\
&\pt_\m \S^{\m,\a\b}=4q^{[\b}u^{\a]}2+\f^{\b\a}.
\end{eqnarray}
To make the aforementioned equation close, we also need the equation of state, which links $e, P, S^{\a\b}$.

In practical use, the aforementioned first-order theory has non-physical modes at the ultraviolet region, which violates the relativistic causality and leads to numerical instability. This problem stems from the constitutive relations \eqs{spin-deco}{spin-deco2} that represent simple proportionality between the responses of the fluid (i.e., LHSs) and the corresponding forces (i.e., RHSs). The simplest means of overcoming this drawback of first-order hydrodynamics is to amend \eqs{spin-deco}{spin-deco2} to the Israel-Stewart form:
\begin{eqnarray}
\label{spin-IS}
&\t_\w (D\s_\w^{\m\n})_\perp+\s_\w^{\m\n}=2\w\pt_\perp^{\lan\m}u^{\n\ran},\\
&\t_\z (D\s_\z^{\m\n})_\perp+\s_\z^{\m\n}=\z\h\D^{\m\n},\\
&\t_\l (Dq^\m)_\perp+q^\m=\l(D u^\m+\b\pt_\perp^\m T-4\O^{\m\n}u_\n),\\
&\t_\g (D\f^{\m\n})_\perp+\f^{\m\n}=2\g(\pt_\perp^{[\m}u^{\n]}+2\O^{\m\n}_\perp),
\end{eqnarray}
where $(\cdots)_\perp$ means taking the components transverse to $u^\m$ (e.g., $(D\s_\w^{\m\n})_\perp=\D^{\m}_\r\D^\n_\s D\s_\w^{\r\s}$). In these equations, $\s_\w^{\m\n}, \s_\z^{\m\n}, q^\m$, and $\f^{\m\n}$ are treated as dynamical variables as well. Therefore, we also need additional initial conditions for them in practical use. They relax to the constitutive relations \eqs{spin-deco}{spin-deco2} after a time scale much greater than the relaxation times $\t_\w, \t_\z, \t_\l$, and $\t_\g$. In this manner, we obtain a set of closed, numerically stable, hydrodynamic equations. The next step is to develop a numerical application to heavy-ion collisions; Hopefully, it can provide us with valuable insights into the spin sign problem.

\section {Chiral and spin kinetic theories}\label{sec:kinetic}
In addition to hydrodynamics, kinetic theory is another commonly used method to study many-body systems in and out of equilibrium. Let us start with a short review of classical kinetic theory.

\subsection{Classical kinetic theory}
Classically, kinetic theory is built based on a single particle distribution function, which is a scalar function defined in the phase space. The physical meaning of the single particle distribution, which we denote as $f(t,\bm{x},\bm{p})$, is the number of particles with a specific space location $\bm{x}$ and momentum $\bm{p}$ at time $t$. The kinetic equation determines the time evolution of $f(t,\bm{x},\bm{p})$ and was first proposed by Boltzmann in the following form:
\ba
\lb\pt_t +\bm{u}\cdot\pt_{\bm{x}}+\bm{F}\cdot \pt_{\bm{p}} \rb f(t,\bm{x},\bm{p})=C(t,\bm{x},\bm{p}),
\label{nrke}
\ea
where $\bm{u}\equiv \bm{p}/m$ is the single particle velocity with particle mass $m$, $\bm{F}$ is the external force, and $C(t,\bm{x},\bm{p})$ is the collision term, which is a functional of $f$. The LHS of the aforementioned equation is the evolution of $f$ due to streaming in the phase space with the existence of the external force field. In other words, the particle at the phase space point $(\bm{x},\bm{p})$ moves with the velocity $\dot{\bx}=\bm{u}$ and the momentum-space velocity $\dot{\bm{p}}=\bm{F}$ at time $t$, which leads to the change in the distribution function $f(t,\bm{x},\bm{p})$. RHS denotes the collision effects among particles that can change the momentum (and possibly also the location) of the particle under study.

Considering the special relativity, we can generalize Eq.~\eqref{nrke} into a relativistic kinetic equation~\cite{DeGroot:1980dk}. We adopt the Minkowski metric $\w_{\m\n}=\diag\{1,-1,-1,-1\}$ and the convention $c=e=k_B=1$, and we define the eight-dimensional phase space coordinates as $(x,p)$, where $x=x^\m=(t,\bm{x})$ and $p=p^\m=(p^0,\bm{p})$, with $p^0$ being the energy coordinate. Particles satisfy the following on-shell condition $p^0=\sqrt{\bm{p}^2-m^2}$. Defining the distribution function $f(x,p)$ in the eight-dimensional phase space, we write the relativistic kinetic equation in the form
\ba
u^\m \pt_\m f(x,p)+ F^\m \pt^p_\m f(x,p)=C(x, p)
\label{rke}
\ea
where $u^\m\equiv p^\m/p^0$ is the single particle four velocity and $F^\m=(F^0, \bm{F})$ is the four external force. The external force is called mechanical if it satisfies the condition $F^\m= \dot{p}^\m $, which leads to the condition $p^\m F_\m =0$ according to the on-shell condition. Furthermore, we obtain $F^0={\bm{F}\cdot \bm{p}}/{p^0}$. In the following discussion, we always assume $F^\m$ to be mechanical. Substituting the solution of $F^0$ into Eq.~\eqref{rke} and using the chain rule
\ba
\frac{\pt p^0}{\pt \bm{p}} \frac{\pt}{\pt p^0} + \frac{\pt}{\pt \bm{p}}\rightarrow\frac{\pt}{\pt \bm{p}}
\ea
we reproduce the form of the LHS of Eq.~\eqref{nrke} in the nonrelativistic kinetic representation.

The relations between physical quantities and the distribution function are readily obtained. The most elementary quantity is the particle density $n(t,\bm{x})$, which is expressed as $n(\bm{x},t)\equiv\int \frac{d^3\bm{p}}{(2\p)^3} f(x,\bm{p})$. The particle three current is defined as $\bm{j}(\bm{x},t)\equiv\int \frac{d^3\bm{p}}{(2\p)^3}\; \bm{u} f(x,\bm{p})$. Combining the particle density and the three current, we obtain the four current as $j^\m(t,x)\equiv(n,\bm{j})$. In relativistic kinetic theory, the covariant four current can be written concisely as follows:
\ba
j^\m(x)=\int \frac{d^4p}{(2\p)^3}\,\d(p^0-\sqrt{\bm{p}^2-m^2})u^\m f(x,p)
\label{rktcurrent}
\ea
where the delta function ensures that the particles are on-shell. Next, we consider the energy-momentum tensor. Classically, the energy-momentum tensor can be explained as the covariant current of the four momentum and thus reads as
\ba
T^{\m\n}(x)= \int \frac{d^4p}{(2\p)^3}\,\d(p^0-\sqrt{\bm{p}^2-m^2})u^\m p^\n f(x,p).
\ea
The energy-momentum tensor is symmetric because the four velocity is proportional to the momentum $u^\m=p^\m/p^0$. The entropy density is defined as $s=-\int \frac{d^3\bm{p}}{(2\p)^3}\, f(x,\bm{p}) \ls \ln f(x,\bm{p}) -1 \rs $. Similarly, we define the covariant entropy current as
\ba
s^\m=-\int \frac{d^4p}{(2\p)^3}\,\d(p^0-\sqrt{\bm{p}^2-m^2}) u^\m f(x, p) \ls \ln f(x, p) -1 \rs.
\ea
The entropy current satisfies the second law of thermodynamics (the Boltzmann H-theorem) $\pt_\m s^\m\geq 0$, where the equality holds in the global equilibrium state.

\subsection{Wigner function in non-relativistic physics}
When quantum mechanics is in action, the aforementioned kinetic theory must be modified. Quantum kinetic theory can be built based on the Wigner function method~\cite{Bonitz1998Quantum}. The Wigner function is the quantum correspondence of the classical distribution function first proposed by Wigner in 1932. In quantum mechanics, the properties of a particle are described by the wave function $\vf(t,\bm{x})$. The dynamics of a non-relativistic particle is governed by the Schr\"odinger equation:
\ba
i\pt_t \vf &=& -\frac{\pt_{\bm{x}}^2}{2m}\vf + V \vf,
\label{schrodinger}
\ea
where $V=V(t,\bm{x})$ is the external potential. After the second quantization, we define the Wigner function as
\ba
W(t,\bm{x},\bm{p})= \int d^3\bm{y}e^{i\bm{p}\cdot\bm{y}}\langle \vf^\ast_+ \vf_- \rangle
\ea
where~$\vf^\ast_+\equiv\vf^\ast(\bm{x}+\frac{\bm{y}}{2},t)$, $\vf_-\equiv\vf(\bm{x}-\frac{\bm{y}}{2},t)$ and $\langle\cdot\cdot\cdot\rangle$ refers to the ensemble average. Note that the Wigner function is real.

The dynamics of the Wigner function is derived from the Schr\"odinger equation \eqref{schrodinger}. Define~$\bm{x}_\pm\equiv\bm{x}\pm\frac{\bm{y}}{2}$. We obtain
\ba
&&\lb\pt_t
+ \frac{1}{m}\bm{p}\cdot \pt_{\bm{x}}\rb W(t,\bm{x},\bm{p})\non
&=&i\int d^3\bm{y}e^{i\bm{p}\cdot\bm{y}}\langle  \ls V(\bm{x}_+,t) - V(\bm{x}_-,t) \rs \vf^\ast_+ \vf_-\rangle,
\label{wig}
\ea
where we have integrated by parts. Next, we suppose that the gradient of the potential $V$ is small so that we can make a gradient expansion. At the first order in $\pt_\bx$, we have $
V(\bm{x}_+,t) - V(\bm{x}_-,t)=\bm{y}\cdot\pt_{\bm{x}}V(\bm{x},t)$ and thus \eq{wig} reduces to
\ba
\pt_t W
+\frac{\bm{p}}{m}\cdot\pt_{\bm{x}}W
-\pt_{\bm{x}}V \cdot \pt_{\bm{p}}W=0.
\label{deowse}
\ea
We thus identify the Wigner function as the single particle distribution function $f(t,\bm{x},\bm{p})=W(t,\bm{x},\bm{p})$ and identify the external force $\bm{F}=-\pt_{\bm{x}}V(\bm{x},t)$. Thus, Eq.~\eqref{deowse} is reduced to the classical kinetic equation~\eqref{nrke} without the collision term. To obtain the collision term, we must start with an interacting theory rather than the Schr\"odinger equation. The Wigner function method is particularly useful in performing the semiclassical approach to the quantum kinetic theory of spinful particles. Therefore, we next discuss quantum kinetic theory as related to spin-$\frac{1}{2}$ particles.

\subsection{Kinetic theory for Spin-$\frac{1}{2}$ fermions}
With the aforementioned warmup preparation, we now consider the Dirac fermions.
We not only introduce the Wigner function for the spinor field~\cite{Vasak:1987um} but also review the derivation of the kinetic theory available in curved spacetime and the external EM field for Dirac fermions~\cite{Winter:1986da,Calzetta:1987bw,Fonarev:1993ht,Liu:2018xip,Liu:2020flb}. In quantum field theory in Minkowski spacetime, the spin-$\frac{1}{2}$ particle is described by the Dirac field $\j(x)$, which is in general a four-component spinor field. We must establish a local flat frame to introduce the spinor into curved spacetime. This is naturally done by using the vierbein field $e^{\m}_{\hat{\a}}$, which can be considered as a coordinate transformation between the general coordinate of the spacetime manifold and the local flat Minkowski coordinate. We use (un)hatted Greek indices to denote local flat (curved) spacetime coordinates. In addition, $\na_\m$ denotes the covariant derivative with respect to the diffeomorphism and $g^{\m\n}$ denotes the curved spacetime metric. The Levi-Civita symbol is $\e^{\m\n\a\b}=\e^{\hat{\m}\hat{\n}\hat{\a}\hat{\b}}/\sqrt{-g(x)}$ with $\e^{\hat{0}\hat{1}\hat{2}\hat{3}}=-\e_{\hat{0}\hat{1}\hat{2}\hat{3}}=1$ and $g=\det{(g_{\m\n})}$.
The dynamics of the Dirac field obey the Dirac equation
\begin{equation}
 [i\hbar\gamma^\mu(\nabla_\mu + iA_\mu/\hbar)-m]\,\psi(x)
 = \bar{\psi}(x)\,[i\hbar({\ola\nabla}_\mu - iA_\mu/\hbar)\gamma^\mu+m]
 = 0\,,
 \label{diraceq}
\end{equation}
where the Dirac matrices satisfy $\lc \g^\m, \g^\n \rc=2g^{\m\n}$, $\nabla_\mu\psi=(\partial_\mu + \Gamma_\mu)\psi$ with the spin connection $\Gamma_\mu= -\frac{i}{4}\sigma^{\alpha\beta} g_{\a\sigma}e^{\hat{\l}}_\b(\partial_\mu e^\sigma_{\hat{\l}} + \Gamma^\sigma_{\mu\nu}e^\nu_{\hat{\l}})$ and the spin matrix $\sigma^{\alpha\beta} = \frac{i}{2}[\gamma^{\alpha},\gamma^{\beta}]$, $A_\m$ is the $U(1)$ gauge potential, and $\bar\psi(x)\equiv \psi^\dag(x)\gamma^\hzero$.

Next, we establish the phase space in curved spacetime to introduce the Wigner function and the kinetic theory. We use the cotangent vector $p_\m$ to denote the momentum in curved spacetime with $y^\m$ as its conjugate variable. Thus, the momentum space is the cotangent space of the spacetime manifold at a given point. The local inner product of the momentum space and the spacetime manifold constitute the phase space, which is the cotangent bundle~\cite{nakahara2003geometry}. $\{y^\m\}$ constitutes the tangent space at a given point of the spacetime manifold, and the tangent bundle is locally the inner product of the tangent space and the spacetime manifold. We introduce the horizontal lifts of the covariant derivative in the cotangent bundle $D_\m = \na_\m + \G^{\l}_{\m\n}p_\l \pt^\n_p$ and the tangent bundle $D_\m = \na_\m - \G^{\l}_{\m\n}y^\n \pt_\l^y$. With the horizontal lift, we can verify that $D_\m p_\n = D_\m y^\n =0$.

The covariant Wigner operator under the U(1) gauge, local Lorentz transformation, and diffeomorphism are defined as~\cite{Liu:2018xip}
\ba
\hat{W}(x,p)=\int\sqrt{-g(x)}d^4y e^{-ip\cdot y/\hbar} \hat{\r}(x,y),
\label{wp}
\ea
with $\hat{\r}(x,y) \equiv \bar{\j}(x,y/2) \otimes \j(x,-y/2)$ and $\j(x,y) \equiv e^{y \cdot D}\j(x)$ , where $D_\m$ also contains the $U(1)$ gauge field when acting on a charged spinor: $D_\m\j(x,y)=(\na_\m-\G^{\l}_{\m\n}y^\n \pt_\l^y+iA_\m/\hbar)\j(x,y)$. The Wigner function is defined by replacing the operator $\hat{\r}(x,y)$ with the ensemble average $\r(x,y)\equiv\langle\hat{\r}(x,y)\rangle$ in Eq.~\eqref{wp}. The dynamics of the Wigner function with full quantum corrections are derived with the help of the Dirac equation \eqref{diraceq}, which can be solved by the expansion method with respect to $\hbar$ with the power counting scheme $p_\m=O(1)$ and $y^\m\sim i\hbar\pt^\m_p=O(\hbar)$~\cite{Liu:2018xip}.
Up to $O(\hbar^2)$, the dynamic equation reads as~\cite{Liu:2018xip}
\begin{equation}
 \begin{split}
   & \ls \gamma^\mu\biggl({\P}_\mu + \frac{i\hbar}{2} \Delta_\mu \biggr)-m \rs W \\
   & = \frac{i\hbar^2}{32}\gamma^\mu
    \biggl(R_{\mu\nu\alpha\beta} +i\frac{\hbar}{6}\partial_p\cdot\nabla R_{\mu\nu\alpha\beta}\biggr)
    \Bigl[\partial_p^\n W,\; \sigma^{\alpha\beta}\Bigr],
 \end{split}
 \label{wignerfms}
\end{equation}
with
\begin{equation}
 \begin{split}
 {\P}_\mu
  &= p_\mu
   	- \frac{\hbar^2}{12}(\nabla_\rho F_{\mu\nu})\partial^\n_p\partial^\rho_p
    + \frac{\hbar^2}{24}{R^\rho}_{\sigma\mu\nu}\partial^\sigma_p\partial_p^\nu p_\rho
  	  +\frac{\hbar^2}{4}R_{\mu\nu}\partial_p^\nu, \\
 \Delta_\mu
  & = \nabla_\mu + (-F_{\mu\lambda}+\G^\n_{\m\l}p_\n)\partial_p^\lambda
  	 - \frac{\hbar^2}{12}(\nabla_\r R_{\m\n})\partial_p^\r\partial_p^\nu \\
  &\quad  - \frac{\hbar^2}{24}(\nabla_\l{R^\rho}_{\sigma\mu\nu})
  	    \partial_p^\nu\partial_p^\s\partial_p^\lambda p_\rho
  	 + \frac{\hbar^2}{8}{R^\rho}_{\sigma\mu\nu}\partial_p^\nu\partial_p^\sigma D_\rho \\
  &\quad + \frac{\hbar^2}{24} ( \nabla_\alpha\nabla_\beta F_{\mu\nu}
  	  + 2{R^\rho}_{\alpha\mu\nu}F_{\beta\rho} )
  	    \partial_p^\nu \partial_p^\alpha\partial_p^\beta,
     \label{ptmd}
 \end{split}
\end{equation}
where $R_{\mu\nu}={R^\rho}_{\mu\rho\nu}$ is the Ricci tensor. We find that the spacetime curvature comes at $O(\hbar^2)$ at least. The Wigner function for the Dirac field is a $4\times 4$ matrix, which is different from the scalar case discussed in the previous subsection. Thus, the relation between the Wigner function and the semiclassical distribution function is less obvious in the spinor case. Equation~\eqref{wignerfms} holds 16 scalar equations if we separate its matrix components, which can be decomposed into hermitian and antihermitian parts further.

Thus, we decompose the Wigner function based on Clifford algebra: $W=\frac{1}{4}[\mathcal{F}+i\g^5\mathcal{P}+\g^\m\mathcal{V}_\m
+\g^5\g^\m\mathcal{A}_\m+\frac{1}{2}\s^{\m\n}\mathcal{S}_{\m\n}]$, where $\gamma^5= (-i/4!)\varepsilon_{\mu\nu\rho\sigma} \gamma^\mu\gamma^\nu\gamma^\rho\gamma^\sigma$ and all the Clifford coefficients are real. Furthermore, Eq.~\eqref{wignerfms} can be decomposed into dynamic equations for the Clifford coefficients. The Clifford coefficients are not independent. We choose the independent variables as $\mc{V}^\m$ and $\mc{A}^\m$. The physical meanings of $\mc{V}^\m$ and $\mc{A}^\m$ are the vector current density and the axial current density in the phase space, respectively, where the latter is also related to the canonical spin current density in phase space. Therefore, the vector current, axial current, and canonical spin current are respectively given by
$J^\m \equiv \langle \bar{\j}\g^\m \j\rangle=\int_p \mc{V}^\m$,
$J_5^\m\equiv\langle\bar{\j}\g^\m\g^5\j\rangle=\int_p \mc{A}^\m$, and
$\ms{S}^{\l,\m\n} \equiv \langle \frac{\hbar}{4} \bar{\j}\{\s^{\m\n},\g^\l\}\j \rangle=-\frac{\hbar}{2}\int_p\e^{\l\m\n\s}\mc{A}_\s$ with $\int_p \equiv\int \frac{d^4 p}{(2\p)^4 \sqrt{-g(x)}}$. In the limit $\hbar\rightarrow 0$, the vector $\mc{V}_\m$ is proportional to the momentum $p^\m$, which is in accordance with Eq.~\eqref{rktcurrent}. However, the axial vector $\mc{A}_\m$ has different forms in the massless and massive cases because spin is parallel (or anti-parallel) to the momentum for a massless particle and is perpendicular to the momentum for a massive particle. Although spin is not an independent variable in the massless case, it induces a Berry curvature, which leads to nontrivial topological effects and results in the chiral kinetic theory. While in the massive case, spin becomes an independent variable, which induces two new degrees of freedom (i.e., the orientation of the spin vector). The complete set of kinetic equations in the massive case is thus composed of four equations. We call this theory the spin kinetic theory. Let us discuss the chiral and spin kinetic theories separately in the following sections.

\subsubsection{Chiral kinetic theory}

For massless fermions, in the classical limit, not only $\mc{V}_\m$ but also $\mc{A}_\m$ is parallel to the momentum, and up to $O(\hbar)$, they read as
\ba
\lb \mc{V},\mc{A}\rb^\m&=&4\p \big\{ \ls p^\m \lb f,f_5 \rb
+\hbar \S_n^{\m\n}\D_\n \lb f_5,f \rb \rs  \d(p^2)\non
&&\quad+ \hbar \widetilde{F}^{\m\n}p_\n \lb f_5,f \rb \d'(p^2)\big\},
\label{masslessaxial}
\ea
where $f=f(x,p)$ and $f_5=f_5(x,p)$ are two scalar coefficients, $\S_n^{\m\n}=\frac{1}{2p\cdot n}\e^{\m\n\r\s}p_\r n_\s$ is the spin tensor for chiral fermions with $n^\m$ being a unit time-like frame vector, and the delta function $\d(p^2)$ imposes the mass-shell condition at a classical limit. Comparing the vector and axial currents for massless fermions with Eq.~\eqref{rktcurrent}, we find that the two scalar functions $f$ and $f_5$ represent the semiclassical vector distribution function and axial distribution function, respectively. The second term in Eq.~\eqref{masslessaxial} is called the side-jump term, which ensures the total angular momentum to be conserved during collisions of two massless fermions~\cite{Chen:2014cla}, whereas the last term comes from the interaction between the spin and external EM field.

We define the right- and left-hand distribution functions as $f_{R/ L}=\frac{1}{2}(f\pm f_5)$. The kinetic equations for $f_{R}$ and $f_L$ are derived as~\cite{Liu:2018xip}
\ba
0&=&\d(p^2\mp\hbar F_{\a\b}\S_n^{\a\b}) \bigg[p_\m \D^\m  f_{R/ L}\non
&&\pm\frac{\hbar}{p\cdot n}\widetilde{F}_{\m\n} n^\m \D^\n f_{R/ L}
\pm \hbar \D^\m \lb \S^n_{\m\n}\D^\n f_{R/ L} \rb\bigg],\quad
\label{keml}
\ea
where the mass-shell condition is corrected by the interaction between spin and the external EM field at $O(\hbar)$.
The flat spacetime version of the aforementioned chiral kinetic equation has been under intensive investigations recently~\cite{Son:2012wh,Stephanov:2012ki,Gao:2012ix,Son:2012zy,Chen:2012ca,Chen:2014cla,Chen:2015gta,Huang:2015mga,Hidaka:2016yjf,Mueller:2017arw,Carignano:2018gqt,Huang:2018wdl,Huang:2018aly,Lin:2019ytz,Lin:2019fqo,Gao:2020ksg}, which can be written in the following form (for right-hand particles only) after $p_0$ being integrated out:
\ba
0&=&\bigg\{\bigg(1-\frac{\hbar(\bm{B}\cdot \bm{p})}{2|\bm{p}|^3}\bigg)\pt_t
+\bigg(\bm{v}-\frac{\hbar}{2|\bm{p}|^3} [(\bm{E}-\na\e_{\bm{p}})\times\bm{p}]\non
&&-\frac{\hbar\bm{B}}{2|\bm{p}|^2}\bigg)\cdot\na
+\bigg((\bm{E}-\na\e_{\bm{p}})
+\bm{v}\times\bm{B}\non
&&-\frac{\hbar}{2|\bm{p}|^3}((\bm{E}-\na\e_{\bm{p}})\cdot\bm{B})\bm{p}\bigg)\cdot\na_{\bm{p}}\bigg\}f_R,
\ea
where we have chosen $n^\m=(1,0,0,0)$, $\e_{\bm{p}}\equiv p_0=|\bm{p}|-\frac{\hbar\bm{B}\cdot \bm{p}}{2|\bm{p}|^2}$ as the particle energy and $\bm{v}\equiv \frac{\pt \e_{\bm{p}}}{\pt \bm{p}}$ as the effective velocity. We find that a phase space correction factor $\lb 1-\hbar\bm{B}\cdot \bm{b}\rb$ exists, where $\bm{b}=\frac{\bm{p}}{2|\bm{p}|^3}$ is the Berry curvature. The dispersion relation is also corrected by the Berry curvature at $O(\hbar)$. The three components for the right-hand particles take the form
\ba
\bm{J}_R&=&\int \frac{d^3\bm{p}}{(2\p)^3}\bigg(\bm{v}-\frac{\hbar \bm{B}}{2|\bm{p}|^2}
-\frac{\hbar}{2|\bm{p}|^3}\bm{E}\times \bm{p}\non
&&+\frac{\hbar}{2|\bm{p}|^3}\e_{\bm{p}} \bm{p}\times\na \bigg)f_R.
\label{cmecinemf}
\ea
Similarly, the kinetic equation and current for left-hand particles can be readily derived.

The kinetic theory in curved spacetime can be used to study the rotating frame. We consider the frame as rotating with the angular velocity $\bm{\O}$ in the inertial frame, and we choose the frame vector $n^\m=(1, x\times\bm{\O})$. The kinetic equation reads as~\cite{Liu:2018xip}
\begin{equation}
 \label{eq:keq2}
 \begin{split}
  & \Biggl[
  (1+2\hbar\,\bm{\O}\cdot\bm{b})\frac{\partial}{\partial t}
  + \Bigl\{\widetilde{\bm{v}}
  + 2\hbar|\bm{p}|(\widetilde{\bm{v}}\cdot\bm{b})\, \bm{\O}
  \Bigr\}\cdot\frac{\partial}{\partial\bx}  \\
  & \qquad
  + 2|\bm{p}|(\widetilde{\bm{v}}\times\bm{\O})\cdot\frac{\partial}{\partial\bm{p}}
  \Biggr] f_R = 0,
 \end{split}
\end{equation}
where~$\widetilde{\bm{v}} = \partial\tilde{\epsilon}_\bp/\partial\bp$ and $
 \tilde \epsilon_\bp = |\bp| - \frac{\hbar}{2}\hat{\bp}\cdot\bm{\O}$.
We find the correspondence between the rotation velocity and magnetic field in $\widetilde{\e}_\bp$ (thus in $\widetilde{\bm{v}}_\bp$) is $|\bp|\bm{\O} \leftrightarrow \bB$, whereas other places is $2|\bp|\bm{\O} \leftrightarrow \bB$. The current is
\begin{equation}
 \label{eq:J_rotating_n}
 \bm{J}_R
  = \int_\bp \biggl[
  \widetilde{\bm{v}}_\bp + 2\hbar|\bp|(\widetilde{\bm{v}}_\bp\cdot\bm{b}_\bp)\,\bm{\O}
 \biggr] f_R + O(\O^2)\,.
\end{equation}

The equilibrium state can be derived from the kinetic equation \eqref{keml}. We suppose the local equilibrium distribution functions depend on a linear combination of the collisional conserved quantities: the particle number, energy and momentum, and angular momentum. Therefore, we have $f^{\rm LE}_{R/L}=n_F(g_{R/L})$ with $g_{R/ L}=p \cdot \b + \a_{R/ L} \pm \hbar\S_n^{\m\n}\o_{\m\n}$, where the coefficients $\b_\m, \a$'s, and $\o_{\m\n}$ depend only on $x$; $\b^\m$ is time-like; and $n_F$ is supposed to be the Fermi--Dirac distribution function. The global equilibrium condition is derived as~\cite{Liu:2018xip}
\ba
\na_\m \b_\n + \na_\n \b_\m &=& \f (x) g_{\m\n},\non
\na_{[\m} \b_{\n]} - 2\o_{\m\n} &=& 0,\non
\na_\m \a_{R/L}&=&F_{\m\n}\b^\n,
\label{equilibriumconml}
\ea
where $\phi(x)$ is an arbitrary function that arises as a result of the conformal invariance in the massless case. We define the four velocity of the fluid as $U^\m\equiv T\b^\m$ with $T$ being the temperature, and the chemical potential $\m_{R/L}\equiv -T\a_{R/L}$. Substituting the global equilibrium condition into Eqs.~\eqref{cmecinemf} and (\ref{eq:J_rotating_n}) and considering also the current of left-hand particles, we derive the CME and CSE as
\ba
\bm{J}=\frac{\hbar\m_5}{2\p^2} \bm{B},\qquad
\bm{J}_5=\frac{\hbar\m}{2\p^2}\bm{B},
\ea
where $ \m=\frac{1}{2}\lb \m_R + \m_L \rb $ and
$ \m_5=\frac{1}{2}\lb \m_R - \m_L \rb $, and the CVEs
\ba
\bm{J}&=&\frac{\hbar}{\p^2}\m\m_5\,\bm{\O},\non
\bm{J}_5^\m&=&\hbar\lb\frac{(\m^2+\m_5^2)}{2\p^2}+\frac{T^2}{6}\rb\bm{\O}.
\ea
We should note that the results for the CME and CVE currents are independent of the choice of the frame vector $n^\m$.

\subsubsection{Spin kinetic theory}

For massive fermions, the particle spin is perpendicular to its momentum up to $O(\hbar)$. The expressions of the vector and the axial vector are as follows:
\ba
\mc{V}^\m
&=& 4\p \bigg\{ p^\m f \d(p^2-m^2)
+m\hbar \widetilde{F}^{\m\n} \h_\n f_A \d'(p^2-m^2)\non
&&+\frac{\hbar}{2m}\epsilon^{\m\n\r\s} p_\n \D_\r \lb \h_\s f_A\rb \d(p^2-m^2)\bigg\},
\label{solutionmsv}
\\
\mc{A}^\m &=& 4\p\big\{m\h^\m f_A \d(p^2-m^2)+ \hbar \widetilde{F}^{\m\n}p_\n f \d'(p^2-m^2)\big\},
\label{mssa}
\ea
where $f=f(x,p)$ and $f_A=f_A(x,p)$ are two scalar functions and $\h^\m$ is the unit spacelike spin vector that is perpendicular to momentum $p^\m \h_\m=0$. We define $f_\pm \equiv \frac{1}{2}\lb f \pm f_A \rb$, which satisfy the following relation:
\ba
4\p m f_{\pm}\d(p^2-m^2\mp \hbar\S_S^{\a\b} F_{\a\b}) &=& \Tr\ls W(x,p) P_\pm(\h) \rs,
\ea
where $\S_S^{\m\n}=\frac{1}{2m}\e^{\m\n\r\s}\h_\r p_\s$ is the spin tensor for massive fermions and $P_\pm(\h)\equiv (1/2)(1\pm\g^5\g_\m\h^\m)$ is the spin projection operator~\cite{kaku:1993ym}. Thus, the physical meanings of $f_\pm$ are the semiclassical distribution functions that describe the spin-up and spin-down states with respect to $\h^\m$. The kinetic equations for $f_\pm$ are derived as~\cite{Liu:2020flb}
\ba
0&=&\d(p^2-m^2\mp \hbar\S_S^{\a\b} F_{\a\b})\non
&&\times\bigg\{\bigg[p^\m \D_\m
\pm\frac{\hbar}{2}\S_S^{\m\n} \lb \na_\r F_{\m\n} -p_\l{R^\l}_{\r\m\n} \rb \pt^\r_p\bigg] f_{\pm}\non
&&+\frac{\hbar}{2}(f_+ - f_-)\biggl[\lb \na_\r F_{\m\n} -p_\l{R^\l}_{\r\m\n}\rb \pt^\r_p\S_S^{\m\n}\non
 &&  - \frac{1}{2m} \tilde{F}^{\n\s} \pt^p_\n \lb p\cdot \D \h_\s -F_{\s\l}\h^\l \rb \biggr]\bigg\}.
\label{keosspud}
\ea
The evolution equation for the spin-direction vector $\h^\m$ is given by~\cite{Liu:2020flb}
\ba
0&=&  \d(p^2-m^2)\bigg[f_Ap\cdot\D \h^\m -
f_AF^{\m\n} \h_\n
+ \h^\m  p\cdot\D f_A \non
&&-\frac{\hbar}{4m }\epsilon^{\m\n\r\a} p_\a \lb \na_\s F_{\n\r} -p_\l{R^\l}_{\s\n\r} \rb \pt^\s_p f\non
&&-\frac{\hbar}{2m} \tilde{F}^{\m\n} \pt^p_\n \lb p\cdot\D f \rb\bigg].
\label{keospin}
\ea
We emphasize that the third term on the right-hand side is actually $O(\hbar)$ order. From the kinetic equations, we can extract the Mathisson--Papapetrou--Dixon equations as
\ba
\frac{Dp^\m}{D\t}&=& F^{\m\l}\frac{p_\l}{m}\pm \frac{\hbar}{2m}\S_S^{\a\b}\lb\nabla^\m F_{\a\b}-p_\l{R^\l}_{\r\a\b}\rb,\\
\frac{D\hbar\S_S^{\m\n}}{D\t}&=&2\frac{1}{m}F_\s^{\;\;[\m}\hbar\S_S^{\n]\s}+2p^{[\m}\frac{dx^{\n]}}{d\t},
\label{eomfors}
\ea
where $\t$ is the proper time along the trajectory of the particle and $d x^\m/d\t=p^\m/m$. The previous two equations describe the spin dynamics for a single particle in curved spacetime and the external EM field.

We can derive the equilibrium state for massive fermions using the same method as in the massless case. Supposing $f^{\rm LE}_{\pm}=n_F(g_{\pm})$ with $g_{\pm}=p \cdot \b + \a_\pm \pm \hbar\S_S^{\m\n}\o_{\m\n}$ and substituting it into Eq.~\eqref{keosspud}, we verify that the following conditions make Eq.~\eqref{keosspud} hold:
\ba
\na_\m \b_\n + \na_\n \b_\m &=& 0,\non
\na_{[\m} \b_{\n]} - 2\o_{\m\n} &=& 0,\non
\a_+ - \a_-&=&O(\hbar),\non
\na_\m \a_{\pm} &=&F_{\m\n}\b^\n,
\label{equilibriumcons}
\ea
where we use $\b^\r \na_\r F_{\m\n}=F_\n^{\;\;\r}\na_\m\b_\r - F_\m^{\;\;\r}\na_\n\b_\r$~\cite{DeGroot:1980dk}. Furthermore, we find that the following solutions of $\a_\pm$ and $\h^\m$ satisfy the spin evolution equation \eqref{keospin}~\cite{Liu:2020flb}:
\ba
\a_+ &=& \a_-,\non
\h^\m&=&-\frac{1}{2m\G}\e^{\m\n\r\s}p_\n\nabla_{[\r}\b_{\s]},
\label{eqspinvec}
\ea
where $\G^2=\frac{1}{2}\na_{[\m}\b_{\n]}\L^{\m\r}\L^{\n\s}\na_{[\r}\b_{\s]}$ with $\L^{\m\n}=g^{\m\n}-\frac{p^\m p^\n}{m^2}$. Thus, the particle spin is polarized along the thermal vorticity at global equilibrium. The spin polarization per particle in the phase space is defined by $S^\m=\mc{A}^\m/(4\p f)$. Substituting the global equilibrium conditions (\ref{equilibriumcons}) and (\ref{eqspinvec}), we obtain
\ba
S^\m_{\rm GE}&=&\frac{\hbar}{4} \e^{\m\n\r\s}p_\n\nabla_{[\r}\b_{\s]}[1-n_F]\d(p^2-m^2)\non
&&+\hbar\widetilde{F}^{\m\n}p_\n\d'(p^2-m^2),
\label{sppmsvpp2}
\ea
which, after integrating $p_0$ over $0$ to $\infty$, yields formula (\ref{spin}) for $s=1/2$ and with the contribution from the EM field added. (Note that in \eq{spin}, the approximation $\sqrt{\bp^2+m^2}\approx m$ is used as $m$ for $\L$ hyperon, for example, is large.) More details on collisionless spin kinetic theory in flat space are provided in Refs.~\cite{Gao:2019znl,Weickgenannt:2019dks,Hattori:2019ahi,Wang:2019moi}. Discussions on the collision terms in CKT and SKT are found in Refs.~\cite{Carignano:2019zsh,Li:2019qkf,Chen:2015gta,Hidaka:2016yjf,Yang:2020hri,Yamamoto:2020zrs}.

\section {Summary}\label{sec:summ}
We discussed some intriguing properties of the strong EM fields and vorticity in heavy-ion collisions. We provided a heuristic introduction to the anomalous chiral transport phenomena and spin polarization in heavy-ion collisions. We briefly reviewed the recent progress in both theory and experiments toward understanding these novel quantum phenomena in heavy-ion collisions. The ACTs could be used to detect the nontrivial topological structure of the QCD gauge sector and the possible P and CP violations of strong interaction in a high-temperature environment. The spin polarization of hadrons provides us a probe to the (local) rotating properties and to the spin dynamics of the quark-gluon matter. This opens a door to a new era of subatomic spintronics.

Some challenges remain. Noticeably, the experimental observables for the ACTs (e.g., the CME) contain strong background contributions, which call for more efforts and new ideas from both the theoretical and experimental sides to be resolved. The experimental data for the azimuthal-angle dependence of spin polarization show a qualitatively opposite trend as compared to the thermal vorticity based on theoretical calculations, which gives rise to a spin sign problem. It is promising that new theoretical frameworks with spin as an independent dynamical variable may provide important insight into the spin sign problem. Presently, two of these frameworks, namely, spin hydrodynamics and spin kinetic theory, are progressing rapidly, and hopefully in the near future, the numerical simulations based on them could be achieved.

\vspace{1cm}

\emph{Acknowledgments}---
We thank J. Bloczynski, W. T. Deng, X. G. Deng, L. L. Gao, K. Hattori, M. Hongo, H. Z. Huang, P. Huovinen, H. Li, Y. Jiang, J. Liao, D. E. Kharzeev, G. L. Ma, Y. G. Ma, K. Mameda, M. Matsuo, L. G. Pang, A. V. Sadofyev, H. Taya, G. Wang, Q. Wang, X. N. Wang, D. X. Wei, H. Z. Wu, X. L. Xia, Y. Yin, S. Zhang, and X. Zhang for collaboration on the topics covered in this article. This work is supported by NSFC under Grant Nos.~11535012 and ~11675041.


\bibliography{ref}

\end{document}